# A composite electrodynamic mechanism to reconcile spatiotemporally resolved exciton transport in quantum dot superlattices

Rongfeng Yuan[1,§], Trevor D. Roberts[1,§], Rafaela M. Brinn[1], Alexander A. Choi[1], Ha H. Park[1], Chang Yan[1,‡], Justin C. Ondry[1,†], Siamak Khorasani[2], David J. Masiello[2,3], Ke Xu[1,4], A. Paul Alivisatos[1,†], Naomi S. Ginsberg[1,4,5,6,7,*]

1. Department of Chemistry, University of California Berkeley, Berkeley, California 94720, United States
2. Department of Materials Science and Engineering, University of Washington, Seattle, Washington 98195, United States
3. Department of Chemistry, University of Washington, Seattle, Washington 98195, United States
4. STROBE, National Science Foundation Science and Technology Center
5. Department of Physics, University of California Berkeley, Berkeley, California 94720, United States
6. Materials Science Division and Molecular Biophysics and Integrated Bioimaging Division, Lawrence Berkeley National Laboratory, Berkeley, California 94720, United States
7. Kavli Energy NanoSciences Institute at Berkeley, Berkeley, California 94720, United States
§ These authors contributed equally to this work
‡ Present Address: Center for Ultrafast Science and Technology, School of Chemistry and Chemical Engineering, Shanghai Jiao Tong University, Shanghai 200240, P. R. China
† Present Address: Department of Chemistry, University of Chicago, Chicago, Illinois 60637, United States
* Corresponding author (nsginsberg@berkeley.edu)

**Abstract**

Quantum dot (QD) solids are promising optoelectronic materials; further advancing their device functionality depends on understanding their energy transport mechanisms. The commonly invoked near-field Förster resonance energy transfer (FRET) theory often underestimates the exciton hopping rate in QD solids, yet no consensus exists on the underlying cause. In response, we use time-resolved ultrafast stimulated emission depletion (TRUSTED) microscopy, an ultrafast transformation of stimulated emission depletion (STED) microscopy to spatiotemporally resolve exciton diffusion in tellurium-doped CdSe-core/CdS-shell QD superlattices. We measure the concomitant time-resolved exciton energy decay due to excitons sampling a heterogeneous energetic landscape within the superlattice. The heterogeneity is quantified by single-particle emission spectroscopy. This powerful multimodal set of observables provides sufficient constraints on a kinetic Monte Carlo simulation of exciton transport to elucidate a composite transport mechanism that includes both near-field FRET and previously-neglected far-field emission/reabsorption contributions. Uncovering this mechanism offers a much-needed unified framework in which to characterize transport in QD solids and additional principles for device design.

**MAIN TEXT**

**Introduction**

Colloidal quantum dots (QDs), semiconductor nanocrystals suspended in solutions due to their surface bound ligand molecules, are highly tunable building blocks for next generation solid-state devices, like displays(*1*), lasers(*2*), or solar cells(*3*). Quantum dot superlattices (QDSLs) are highly ordered arrangements of QDs (**Figure 1a**) that can enable bottom-up design of hierarchically organized functional materials(*4–6*). Generally, SLs are desirable over disordered dropcast films of QDs for their high spatial order and superior transport properties. The primary photoexcited species in QDs is an exciton - an electron-hole pair that resides on a QD due to confinement effects. Exciton transport is therefore an important process for the function of QD-based optoelectronic solid devices. Understanding exciton transport mechanisms is a key step towards developing design principles that enable finer control over exciton transport. For example, QD-based solar cells benefit from high exciton diffusivity because these photogenerated excitations must travel from the site of absorption to a charge separation interface(*7*); in displays, little-to-no exciton diffusivity is desirable because exciton transport among QDs of different emitting colors or to quenching sites will cause color accuracy and low photoluminescence yield issues(*8*).

Exciton transport in QD solids has increasingly been studied via spatiotemporally resolved photoluminescence(*9, 10*) and transient absorption(*11, 12*) in order to determine the relationship between transport and microstructure and to reveal the nature, e.g., normal or anomalous diffusion, of transport. Förster resonance energy transfer (FRET) is the most commonly invoked model for exciton transport in QD solids(*9, 13–15*) in which the ligand coat is long (e.g., ~2 nm) and insulating, as the center-to-center distances between QDs are therefore large. Yet, FRET theory often underestimates exciton diffusivity compared to experimentally derived values(*9, 11, 16, 17*). Many explanations have been proposed, yet no conclusive cause has been determined. Separately, photon emission and reabsorption has been recently demonstrated to be important for exciton transport in optoelectronics. For example, repeated emission and reabsorption due to total internal reflection, i.e., waveguiding, often referred to as photon recycling, has enhanced transport in metal halide perovskite thin films and nanocrystal solids(*18, 19*). Its potential impact in other materials, including in the well-studied colloidal II-VI semiconductor QD solids, is relatively unexplored. While both mechanisms involve coupling of donor and acceptor transition dipole moments (TDMs), the distance scalings of their respective exciton transfer rates are quite different. FRET, a near-field phenomenon, scales as $r^{-6}$, while emission and reabsorption occur in the far-field, scaling as $r^{-2}$. Discerning these two mechanisms with spatiotemporal microscopy alone is, however, challenging because additional constraints are needed to properly specify a model that includes them both.

The evolution of the exciton energy could generate the additional needed constraints. Both near-field and far-field mechanisms require energetic resonance between a donor and an acceptor QD. Better overlap of the donor emission spectrum and acceptor absorption spectrum leads to a higher probability of exciton transfer. Due to the presence of a Stokes shift, a small energy relaxation of an electronic excitation, the bias of transferring an excitation from a donor to a lower-energy acceptor, and thus a corresponding decay of the average exciton energy are present in both mechanisms. This bias leads to time dependence and correlation of the exciton energy and diffusivity during transport, especially because the bandgap energies of QDs in a solid are generally distributed over a range of values, restricting the number and proximity of potential acceptors of a given donor QD's exciton. Exciton diffusivity is therefore reflected not only in spatiotemporally resolved microscopy but also in measurements of the decay of the exciton energy as a function of time mediated by either near-field or far-field mechanisms. Importantly, however, the roles of energetic resonance and distance scaling in determining the energy transfer rate through these two different mechanisms differ. The donor-acceptor distance plays a much more important role in FRET than in emission/reabsorption. Intuitively, given a specific amount of energy loss over the exciton lifetime, emission/reabsorption should generate more substantial transport because finding a resonant acceptor QD is not as challenging in a heterogeneous energy landscape as it is in FRET because there are no severe distance constraints. Thus, measuring both the exciton mean energy and diffusivity as a function of time(*9*) should allow the relative contributions of the two mechanisms to be quantified and distinguished.

Here, we indeed elucidate mixed-mechanism excitonic energy transport by combining time-dependent exciton energy and diffusivity measurements in a heterogeneous QDSL energy landscape, leveraging a multimodal background-free fluorescence approach. We find that FRET and emission/reabsorption together explain exciton transport behavior in highly ordered Te-doped CdSe/CdS core-shell QDSL monolayers, with a smaller far-field contribution providing a substantial increase to the overall transport. To do so, we first quantitatively characterize the heterogeneous energetic landscape of CdSe:Te/CdS QDSL monolayers by extracting inhomogeneous and intrinsic spectral components of QDs by single-particle emission spectroscopy(*20, 21*). We next track the time-dependent decay of the mean exciton energy by time-resolved emission spectra (TRES) after QDSL photoexcitation. Finally, we measure exciton transport by monitoring the spatiotemporal expansion of an exciton population after a local photoexcitation in QDSL monolayers with time-resolved ultrafast stimulated emission depletion (TRUSTED) microscopy(*22*) and use all three types of experimental results to constrain parameters in a kinetic Monte Carlo (KMC) simulation. Our findings reveal that often overlooked far-field dipole-dipole coupling is an important energy transfer mechanism to include for colloidal QD solids. Furthermore, the relatively high reported exciton migration length demonstrates the possibility of using doped QDs to decouple energetics from QD size in optoelectronic devices.

**Results**

To discern FRET and emission/reabsorption in QDSLs, we focus on 5% Te-doped CdSe (core) / CdS (shell) QDs (diameter 6-7 nm) with mixed oleic acid and oleylamine ligands whose shape resembles a short hexagonal prism, like the nut of a nut & bolt (**Fig. 1A,B**). Importantly this nanocrystal shape allows for us to achieve highly ordered domains of self-assembled nanocrystals where the nanocrystals have well defined interparticle spacing, crystallographic orientation, and minimal structural disorder(*23*). These readily hexagonally pack into an ordered superlattice. Sample preparation and characterization are described in **Supplementary Materials (SM) sections 1 and 2**. The primary effect of Te doping is to broaden the QD emission spectra(*24*) as shown in **Fig. 1C (red curve)**, in contrast to the undoped control (blue curve). Single-particle emission spectroscopy of the 5% Te-doped CdSe/CdS QDs in comparison to their undoped CdSe/CdS control QDs shows that both the distribution of single QD emission peak energies (**Fig. 1D**) and spectral widths (**Fig. 1E**) increase substantially with doping (see **SM Section 1**, **fig. S1**), suggesting that the increased emission bandwidth observed in bulk solution in **Fig. 1C** is due to broadening on both the single QD and ensemble levels. After forming a QDSL monolayer (**SM Section 2, figs. S2-S6**), the emission peak redshifts relative to the solution emission.

To track the time-dependent evolution of the average exciton energy and migration (schematics in **Fig. 2A**), both of which are needed to determine the relative contribution of near- and far-field dipole-dipole coupling in the QDSL monolayer, we employ TRES and TRUSTED, respectively. TRES of the CdSe:Te/CdS QDSL are shown in **Fig. 2B**, and the peak energy emission red-shift is evident as the time delay increases. The associated time resolved mean energy relaxation is shown in **Fig. 2C** for the doped QDs in solution (**pink**) and monolayer QDSL (**red**) along with the corresponding controls for undoped QDs (**light and dark blue**, respectively). The energy loss plotted corresponds to the difference between the average emission energy at each recorded time delay following ~ 100 ps photoexcitation at 470 nm and the peak emission energy at zero time delay. QDs have been demonstrated to be amenable to stimulated emission depletion (STED) super-resolution fluorescence microscopy(*25, 26*). Here, we use its ultrafast adaptation, TRUSTED. To characterize exciton migration in the QDSL monolayer, we report the fraction of excitons remaining at a given time delay after a sub-diffraction population is prepared within the monolayer in a TRUSTED experiment (**Fig. 3A,B**). The initially prepared population of excitons consists of those generated by a few-ps diffraction-limited 550 nm pump pulse (**blue**, **Fig. 3A**) that are not quenched by a nearly coincident 100-ps 740 nm annular 'STED' pulse (**yellow**, **Fig. 3A**). The remaining excitons at a given time delay correspond to those that are not quenched by an identical annular STED pulse at that time delay (see **Methods**, and **Ref.** (*22*) for further details). The net result is that the data in **Fig. 3B**) may be fitted to a simple model of the TRUSTED protocol(*22*) (red curve, **Fig. 3B**) to determine migration parameters, such as diffusivity. At present, TRUSTED time delays of up to ~ 5 ns can be recorded with ~ 100 ps time resolution, making TRES's much longer time delay window highly complementary. The TRES and TRUSTED data, along with single-particle and bulk emission spectroscopy, are used to constrain a KMC simulation of exciton transport that tracks both energy and location of excitons schematically represented in **Fig. 2A**. Below, we first describe the results obtained from the two experimental methods and then describe how these are analyzed in combination with the simulations.

We focus first on the TRES results in Fig. **2C**. For both solution phase QDs and QDSLs, we focus our comparison of the decay and characteristic decay time within the first 15 ns out of the full window of 150 ns measured (**fig. S7-S8**), when the signal-to-noise ratio is highest. As shown in **Fig. 2C**, a more pronounced dynamic redshift in the QDSL (9.8 meV for the undoped (dark blue) and 80 meV for the doped (red)) is observed from TRES compared to from QD solutions (~ 0 for the undoped in **light blue**, and 15 meV for the doped in **pink**). In addition, the Te-doped samples have a greater magnitude of redshift over the exciton lifetime than their undoped counterparts. The enhanced dynamic redshift occurring in the QDSLs relative to the solution phase QDs immediately following excitation indicates the existence of non-equilibrium exciton transport. In other words, the average exciton energy initially decreases as a function of time. This observation means that the additional redshift in the SL state is not a property of isolated QDs but a result from the interaction among QDs. The large redshift in the doped QDSL suggests that its energetic landscape is very heterogeneous (**Fig. 1C**), in qualitative agreement with single-particle emission spectroscopy results (**Fig. 1D**), in addition to the breadth of the constituent QD energies also obtained from single-particle measurement (**Fig. 1E**). Similarly, the undoped QDSL exhibited a smaller amplitude of energy relaxation (dark blue in **Fig. 2C**), in accordance with the narrower energy distribution in **Fig. 1D** (blue).

We use TRUSTED to spatially resolve exciton migration within the CdSe:Te/CdS QDSL monolayers, whose broadened and redshifted emission relative to the absorption onset are ideal for STED microscopy. The undoped CdSe/CdS QDSL system has too small of a Stokes shift to be amenable to STED microscopy or TRUSTED. We fit our data in **Fig. 3B** with a time-dependent diffusivity model (red curve). The initial diffusivity is $(4.6 \pm 2.5) \times 10^{-3}$ cm$^2$/s, and a roughly four-fold decrease in the exciton diffusivity occurs within the 5-ns measurement window. In a TRUSTED dataset, the horizontal axis is the time delay between the first and second STED pulses. The vertical axis corresponds to the fraction of remaining excitons, calculated as the ratio of the photoluminescence with and without the second STED pulse and then normalized to its initial value to construct the "fraction of remaining excitations" in the detection volume. A decaying signal means that excitons move outward between the time delay of the two STED pulses, and the slope of the decay depends on the rate at which excitons travel. In the simplest model, excitons are assumed to undergo a random walk, and the aggregated behavior of these excitons is similar to Brownian motion. A time-dependent diffusivity fit to the TRUSTED data is warranted, given TRES reveals that the mean exciton energy decreases with time. As an exciton proceeds to sample lower energy sites, the number of acceptor sites concomitantly decreases, which would lead to a time-dependent reduction in the energy transfer rate. By approximating this decay with an exponentially decaying exciton diffusivity model using $D(t) = D_o \exp(-kt) + D_c$, the fit diffusivity at time $t=0$ is $D_o = (4.6 \pm 2.5) \times 10^{-3}$ cm$^2$/s, the decay rate $k = (0.0008 \pm 0.0010)$ ps$^{-1}$, and the long-time diffusivity $D_c = (1.5 \pm 1.2) \times 10^{-3}$ cm$^2$/s. Due to the limited signal-to-noise ratio, the error bars on these cited diffusivity parameters are somewhat substantial. Nevertheless, the uncertainty in the characteristic extent of exciton transport within the TRUSTED measurement window, $\sqrt{\int D(t) \cdot dt}$ is small: $35 \pm 6$ nm. (The extrapolated characteristic extent of exciton transport over the QD lifetime, typically referred to as $L_D$, is ~47 nm.) Various control experiments to ensure that the decay of the TRUSTED signal originates from exciton transport are provided in **SM Section 4, figs. S9 and S10**. In addition, measurements of 2 and 7% CdSe:Te/CdS QDSL monolayers can be found in **SM Section 4, fig. S11 to S14**. They yield exciton diffusivities of the same order of magnitude as the 5% Te-doped sample. (The synthesis of these nanocrystals occurred in a different batch from the 5% doped ones. To avoid convolving the associated potential differences in QD properties with the differences in doping concentrations, we refrain from making meaningful comparisons between the 5% batch and the 2% and 7% batch.)

In order to best relate the above energy relaxation and energy migration dynamics, we first characterized the 5% CdSe:Te/CdS QD's heterogeneous energy landscape with single-particle emission spectroscopy. As shown in **Fig. 1D**, compared to undoped CdSe/CdS QDs, the 5% CdSe:Te/CdS emission peak distribution is much wider, indicating a more heterogeneous distribution of QD bandgap energies. The heterogeneity is most likely a result of variability in Te doping amongst individual nanocrystals, like the relative number and/or location of dopants within a given QD, as opposed to size variability within the ensemble. Franzl et al.(*27*) demonstrated two distinct subpopulations of Te dopants in CdSe/ZnS QDs probed via single-particle emission spectroscopy. One subpopulation had one to a few Te dopants and a similar linewidth to undoped particles, and another subpopulation had many Te dopants and much broader spectra. Our single-particle emission results suggest that at 5% doping, our QDs likely only show the "many dopants" regime proposed by Franzl *et al*.(*27*), as we do not observe a distinct population of "few dopants." Given differences in our synthesis procedure such as higher temperature, it would not be surprising that our spectroscopy differs from the much earlier work of Franzl et al., likely resulting from higher and more homogeneous Te incorporation within the CdSe core.

Having established the nature of the energy landscape of the QDSL monolayers, we can not only assert that TRES in QD monolayers reveals the decay of mean exciton energy in QDSL monolayers caused by exciton transport (**Fig. 2A and 2C**), but that the rates of mean energy decay and exciton transport are proportional to one another. Since the mean energy decay rate $k_{\Delta E}$ in the undoped QDSL (**Fig. 2C**, dark blue) is larger than that of the doped QDSL (**Fig. 2C**, red), exciton transport in the undoped QDSL appears to be faster than the doped case. We deduce that the extra redshift in the SL state is not a property of the isolated QDs (**Fig. 2C**, pink), but a result of the interaction among QDs. A spectral shift absent in QD solution but present in QD solids is often taken as evidence for energy transport(*9, 16, 28*) facilitated by the QD proximity in solid films. During this energy transport, excitons on average sample progressively lower energy QDs as they explore the spatioenergetic landscape (**Fig. 2A**). The TRES decay amplitude is therefore a measure of energy heterogeneity, and the decay rate is proportional to the exciton transport rate. Furthermore, the single-particle emission spectroscopy and the TRES measurement of the QD energy heterogeneity support the picture of exciton hopping occurring from high-to-low energy sites on a heterogeneous energetic landscape. A single-exponential fit to the energetic relaxation of the form $\Delta E = \Delta E_\infty [1 - \exp(-k_{\Delta E} t)]$ is performed, where $\Delta E_\infty = -\frac{\sigma_{\mathrm{ih}}^2}{k_\mathrm{B} T}$, $\sigma_{\mathrm{ih}}$ is the width of the inhomogeneous distribution of site energies, and $k_\mathrm{B} T$ is the thermal energy at room temperature (26 meV)(*29*). Based on this fit, $k_{\Delta E} = 0.045$ ns$^{-1}$ and $\sigma_{\mathrm{ih}} = 56$ meV for the doped QDSL, a bit smaller than the result from single-particle emission spectroscopy, which is 74 meV after converting from FWHM in **Fig. 1C**. Using the ~± 20 meV spread in the measured TRES energy loss at long times (**fig. S8**) as a bound for uncertainty of $\sigma_{ih}$, this value agrees well with the single-particle emission spectroscopy.

**Discussion**

We have presented two measurements that inform on exciton transport in the doped QDSLs – TRES, monitoring the rate of the mean energy change, and TRUSTED direct spatiotemporal measurements of the expanding exciton spatial distribution. Although FRET is the most commonly invoked mechanism to describe energy transfer between QDs with long alkyl ligands(*9*, *16*, *30*, *31*) we find that it is inadequate to justify the experimentally observed exciton diffusivity and to reconcile the TRES and TRUSTED results: FRET predicts an exciton diffusivity 100-1000 times smaller than the diffusivity reported by TRUSTED (**SM Section 5**); furthermore, an energy relaxation rate consistent with the TRUSTED observation, according to a generously parametrized FRET model, is much faster than the rate obtained by TRES. In fact, the canonical $R_o$ parameter, the distance at which energy transfer is 50% efficient, would have to be unphysically high to recapitulate the observed diffusivity. This discrepancy has been reported previously by others, and additional considerations investigated include the degree to which a point-dipole approximation is valid, the potential contribution of higher-order dipole-quadrupole interactions, the potential for more optimal dipole-dipole orientation in a solid assembly, and the potential for a larger absorption cross section in films than what is estimated solution(*16*, *17*, *32*) To illustrate the incompatibility of the TRUSTED and TRES results within a FRET framework, we show in **Fig. 2C** that if $R_o$ is adjusted to match the exciton diffusivity extracted by TRUSTED, then it predicts a much faster decay of the mean exciton energy than TRES reports (gray curve in **Fig. 2C**). We find that our two measurements can, however, be reconciled by including contributions from higher order terms in the dipole-dipole coupling expansion elaborated below from which FRET retains only the steep $r^{-6}$ near-field term (black curve in **Fig. 2C**). With $r^{-6}$ dependence alone, exciton transfer is largely limited to 1 to 2 shells of nearest neighbor QDs, and the spectral overlap (requirement of exciton energy resonance) between QDs has less effect on the transfer rate than the distance between QDs. Yet, alleviating the distance penalty for the exciton transfer by introducing higher order (far-field emission/reabsorption) terms allows hops well beyond the nearest neighbors that enjoy better energy resonance, just as sketched in **Fig. 4**. As we show below, even very few of these hops result in a much higher effective diffusivity, where the decay in mean exciton energy remains slow enough to match TRES.

To arrive at this conclusion, we developed a KMC simulation of energy transport by dipole-dipole coupling of QD TDMs, in which the intermediate ($r^{-4}$) and far field ($r^{-2}$) terms of the dipole-dipole interaction expansion can be included. As explored by Andrews *et al.*(*33*, *34*), the energy transfer rate between donor TDM and acceptor TDM can most generally be written as

$$k_{DA} = (k_{near} + k_{intermed} + k_{far})\exp(-\alpha r), \qquad \text{Eq (1)}$$

where $1/\alpha$ is the attenuation length of photons absorbed when traveling inside the medium of donors and acceptors, set here to 400 nm, based on measurement(*35*), and $k_{near} = \frac{1}{\tau_D}\frac{R_o^6}{r^6}$ as already described. The terms $k_{\text{intermed}}$ and $k_{\text{far}}$ also both depend on $\frac{1}{\tau_D}$, the reciprocal of the exciton lifetime, but scale with $r^{-4}$ and $r^{-2}$, respectively, with $r$ being the dipole-dipole distance. Whereas the standard FRET $r^{-6}$ term dominates in the near-field in which the donor-acceptor distance is much smaller than the reduced wavelength $\lambda/2\pi$, the far-field $r^{-2}$ term dominates when $r$ is much greater than $\lambda/2\pi$, and the intermediate $r^{-4}$ term is important when the donor-acceptor separation is comparable to $\lambda/2\pi$.

Importantly, a major contribution to $R_0^6$ is the overlap integral between acceptor absorption and donor emission spectra, which is weighted by a function of the optical frequency $\omega^{-4}$ (or, $\lambda^4$): $\int \sigma_A(\omega)\overline{F_D(\omega)}\omega^{-4}d\omega$. (Here, $\sigma_A(\omega)$ is the wavelength-dependent absorptivity of the acceptor chromophore species, and $\overline{F_D(\omega)}$ is the fluorescence emission spectrum normalized by its area; see **SM Sections 5 and 6**.) The corresponding spectral overlap integral for the far-field contribution is simply $\int \sigma_A(\omega)\overline{F_D(\omega)}d\omega$ without any spectral weighting, and the intermediate term is weighted by $\omega^{-2}$. Relative to the near-field term, the scalings of these weightings effectively enhance the higher order terms 4- and 17-fold, respectively. In our KMC simulation, we therefore embodied **Eq 1** by introducing correlated unitless coefficients for the two higher-order terms,

$$k_{\text{DA}} = \frac{1}{\tau_D}\frac{R_0^6}{r^6}(1 + c_{\text{intermed}}(\frac{2\pi}{\lambda})^2 r^2 + c_{\text{far}}(\frac{2\pi}{\lambda})^4 r^4)\exp(-\alpha r), \quad \text{Eq (2)}$$

in which $c_{\text{intermed}} \cong \sqrt{c_{\text{far}}}$ to reflect these differences. While we expect boundary conditions associated with the field and medium geometry to also play a role (see below), these coefficients introduce a relative importance between terms that should be sufficient to reflect **Eq 1** in the KMC simulation.

When $c_{\text{far}} \sim 27$, it is possible to employ the same model parameters to simultaneously fit both the TRES decay (**Fig. 2C, black**) and the exciton diffusivity extracted via the TRUSTED measurements in **Fig. 3B**, and the value returned for $R_0$ is a very reasonable 7.6 nm. Full details of the simulation can be found in the **SM Section 6**, including an exploration of the role of the different terms in the expansion that demonstrates more explicitly how neither near- nor far-field term alone is able to simultaneously recapitulate the TRUSTED energy transport and TRES energy decay (**Fig. S15**). This exploration further supports that a small amount of emission/reabsorption (far-field interaction) is consistent with our data and suggests that a similar interpretation may explain other spatiotemporal transport studies in QD solids and potentially other systems. Our hypothesis is that both near-field and far-field coupling occur. The near-field coupling predominantly supports exciton hopping to nearest-neighbors of the originally excited QD (**see SM Section 6**), and on its own would require $R_0$=15 nm, beyond acceptable FRET radii. The modest 6% fraction of hops associated with the far-field coupling (**SM Section 6**) enable an overall measured exciton migration length of ~35 nm within the first 5 ns. Far-field effects have been demonstrated in many other systems, including perovskite thin films(*36*). Whereas thin films are sufficiently thick to support traditional waveguiding and enhance the far-field contribution via photon recycling, no traditional waveguide modes can be supported in materials as thin as QDSL monolayers, and more complex mechanisms are needed to explain the relative enhancement of far-field over near-field coupling contributions.

The electrodynamic treatment of the donor-acceptor transfer rate in **Eqs. 1 or 2**, accounts for a substantial portion of $c_\text{far} \sim 27$. Namely, the combination of the ratio of different overlap integrals of the near- and far-field terms and different orientational factors (see **SM Section 6**) accounts for all but a factor of ~5. There are multiple additional potential contributors to this stronger relative far-field contribution that satisfies the simultaneous constraints imposed by the TRUSTED and TRES experimental results. First, **Eqs. 1 and 2** assume isotropic orientational distributions of TDMs, which do not necessarily accurately reflect the weighting of donor TDM orientations in the QDSL monolayer. One primary contribution to reweighting the TDM distribution is that the dielectric boundary conditions can strongly impact the radiative rate when the donor and acceptor are confined to a narrow plane of dielectric constant greater than the surrounding media, as is presently the case. Consider the QDSL monolayer as a nanoconfined cavity, i.e., a 9-nm thick slab whose dielectric constant exceeds those of its surroundings. Emission from TDMs oriented parallel to the monolayer plane should be entirely suppressed in the present limit that the cavity thickness is much lesser than the emission wavelength. Nevertheless, TDMs perpendicular to the plane retain non-zero emission amplitude(*37, 38*) and can furthermore be enhanced due to the Purcell effect – an alteration of the radiative rate due to proximity to a dielectric object(*39*). Typical dipole radiation patterns of TDMs oriented normal to the plane favor emission in the plane via a $\sin^2\theta$ dependence, an effect that is not explicitly accounted for in **Eq. 2** but that can indeed be encapsulated in $c_\text{far}$, and which is distinct from the orientational contributions to $R_o^6$. Furthermore, the emission direction probability can be substantially modified based on the specific cavity thickness and the dielectric mismatch between the cavity and the surroundings, with strong in-plane enhancements(*40, 41*). Although a full quantum electrodynamics treatment with the appropriate boundary conditions for this system is beyond the scope of the current work, the combination of radiative enhancement and in-plane directional bias provides another explanation for the apparent relative strength of the far-field contribution to the donor-acceptor transfer rate in **Eq. 2**. We solidify this claim through simulations of the enhancement of far-field to near-field strengths in the plane of an explicit lattice of TDMs in the presence of a dielectric medium in **SM section 7** and **fig. S16**. Finally, a number of recent reports describe additional plasmonic, polaritonic, or evanescent waveguiding transport mechanisms(*42–44*), which could additionally contribute to the stronger far-field contribution consistent with our observations. Regardless, from our data presented herein, we find that only by including interactions via far-field dipole-dipole coupling do we obtain a model capable of reconciling the two independent measurements of energy transport, TRES and TRUSTED, in CdSe:Te/CdS QDSL monolayers, and we presented multiple possible explanations for the model parameters thus obtained, the most substantial of which fall directly out of the electrodynamic equations.

This work provides a more general treatment of dipole-dipole coupling mediated energy transfer in electronically coupled systems that goes beyond near-field interactions. Prompted by explaining our exciton transport observation in colloidal QD solids, subject to importantly included spectroscopic constraints, for a heterogeneous energetic landscape, we have shown the value of tracking and correlating the energetic relaxation and spatial expansion of excitons in the system under study to distinguish different energy transport pathways. Different materials that support exciton transport may differ in the relative balance of near-field (FRET) and far-field (emission/reabsorption) energy transfer events exhibited. To enhance long-range exciton transport, the two mechanisms call for different optimization strategies. For example, to enhance the far-field mechanism, which can offer strong advantages in increasing the efficacy of an optoelectronic device in which the energy transport material must direct as many excitons as possible to a specific location to transduce their energy into useful work, decreasing the concentration of chromophores could be helpful, so that photons emitted in-plane can travel further before reabsorption and to suppress FRET. In addition, designing materials with high dielectric constants or having the exciton host-material sandwiched by high dielectric constant materials could enhance far-field coupling through Purcell-like effects. Alternatively, when near-field coupling easily dominates the far-field coupling due to the distance scaling, excitons can be better spatially contained, which is helpful in display applications. By decreasing Purcell-like effects through index matching of materials surrounding the active layer, far-field coupling could be deliberately reduced. Finally, to optimize long-range near-field/FRET-like transport, achieving a more homogeneous site energetic landscape is clearly important to prevent exciton trapping at a low energy site, but in the absence of this homogeneity, far-field contributions can compensate.

We noted earlier that other spatiotemporal measurements of exciton transport in QD solids have generated results that suggest transport in excess of what is predicted by FRET theory. We propose that incorporating the possibility of higher order terms in the dipole-dipole coupling of donor and acceptor TDMs could reconcile not only our findings but also those of others that have remained at odds with FRET. For example, Akselrod *et al*(*9*)., who measured both spatial and spectral diffusion via time-resolved photoluminescence, found a characteristic migration length of 32 nm, which would require $R_o$ to be 11 nm, although the expected $R_o$ based on steady state properties is around 5.5 nm; Mork *et al*.(*16*) found that the experimental $R_o$ between a donor and an acceptor QD is ~ 9 nm while the estimation based on the classic formula is ~ 5 nm and explore various possible explanations without arriving at a specific consensus. Applying our model to a specific example from Akselrod *et al*. we find that their measured migration length would include a modest amount of emission/reabsorption with $c_{far}$ ~ 35 very comparable to our own, and a decrease of $R_o$ to a more acceptable value of 5 nm (**SM Section 8**). Further investigating related patterns from other measurements may provide a much-needed unified framework in which to characterize QD solids and also with which to leverage different device design principles.

In conclusion, we present a comprehensive study of the exciton transport mechanism in a CdSe:Te/CdSe QDSL monolayer that reconciles larger-than-expected exciton diffusivities. To do so, we measured the spatiotemporal dynamics of exciton diffusivity via TRUSTED and the accompanying spectrotemporal mean energy relaxation dynamics via TRES. Combining these with explicitly measured QD energy heterogeneity extracted from single-particle measurements, we deduced that the apparent diffusivity must be an average of near- and far-field transport. Such a conclusion would not be possible without all three measurements, since these leverage spectroscopic, in addition to spatial and temporal observables. Together, these different datasets provide sufficient constraints to supply to a KMC simulation model of exciton transport to elucidate the composite transport mechanism, which would not have been possible with only a subset of the techniques used. With this model, we infer that far-field interactions should even more dramatically enhance exciton transport when the energy inhomogeneity is lower (see **SM section 9**, **Table S2**). This powerful multimodal approach is generally applicable and can be readily applied not only to other QD solids but to other optoelectronic materials. The recognition of far-field contributions to exciton transport by including them in a transport model also promises to better explain spatiotemporal transport measurements more broadly, resolving a long-standing paradox presented by the use of FRET-exclusive models. Incorporating this model directly into optoelectronics device design will furthermore assist in generating more optimal functions.

**Materials and Methods**

**Sample preparation**: Te doped CdSe/CdS QD with oleic acids and oleylamine ligands were synthesized with further details in the **SM**. They self-assemble into a QDSL at liquid-air interfaces and are then transferred to a glass substrate. The sample is then encapsulated in a nitrogen atmosphere with UV-cured epoxy before optical characterization.

**TRUSTED**: TRUSTED(*22*) is employed in a home-built confocal microscope with a 63× 1.4NA Plan Apo Leica objective (HC PL APO 63×/1.40 oil CS2, Leica Material #11506350). The excitation and depletion laser pulse trains at 200 kHz were derived from third-harmonic and second-harmonic noncollinear optical parametric amplifiers (NOPA) (Light Conversion), respectively, pumped by a 10 W Light Conversion PHAROS regeneratively amplified laser system with a fundamental wavelength of 1,030 nm. The 25-fJ excitation pulse was centered at 550 nm with 20 nm bandwidth, and the two 125-pJ depletion pulses were centered at 740 nm with a bandwidth set to 16 nm. Both the pump and depletion (STED) pulses were fiber coupled into single-mode polarization-maintaining fibers to produce high-quality Gaussian modes. A vortex phase mask (RPC Photonics VPP-1a) was then used to generate the annular depletion pulse beam mode. The pulses were then directed through a quarter waveplate positioned to circularly polarize the depletion pulses. During the experiment, the sample is rastered with a PI Nano scanning piezoelectric stage (P-545.3C7) in steps of 15 μm over a 60 μm × 60 μm area. Data from the resulting twenty-five spatial locations are averaged to improve the signal-to-noise ratio. Epifluorescence is collected between 687.5 and 712.5 nm through dichroic mirrors and emission filters (three 700/25 filter from Edmund Optics) and is focused onto a fast-gated SPAD detector with a 200-ps rise time (A. Tosi, SPAD lab, Politecnico di Milano; PicoQuant) controlled by a Picosecond Delayer (MPD) that is triggered just after the arrival of the second depletion pulse to eliminate fluorescence occurring before the definition of the detection volume. We phase lock the detection data stream to the timing of an optical chopper (Newport 3501) placed in the excitation pulse line, so that we may separately determine the photon count rates during the 'excitation on' and 'excitation off' chopper phases for multiple cycles. The count rates obtained during these open and closed phases of the chopper are each corrected for the classic pile-up effect with a simple Poisson correction factor(*45*) before we take the difference of the two to isolate the count rate that is attributed to the modulated excitation pulse only. The second STED pulse is separately modulated with a shutter so that data collected when this pulse is blocked can be used as a reference and control. The signal versus delay time obtained when this second depletion pulse is unblocked is divided by the signal versus delay time observed when it is blocked. The resulting data is then normalized to the extrapolated value of this ratio at zero delay time to calculate the fraction of remaining excitations in the detection volume as a function of the delay time.

**TRES**: The sample was pumped by a 470 nm (Picoquant diode laser) laser with ~ 100 ps temporal resolution. Photoluminescence spectra from 550 nm to 750 nm were collected in a Picoquant FluoTime 300 Fluorometer at normal incidence with wavelength step size of 2.5 nm for the CdSe:Te/CdS QDSL sample. Luminesced photons were collected by a PMT detector, and the factory-set wavelength-dependent response of the PMT was used as a calibration curve to correct the individual spectra. The peak emission wavelength was extracted by fitting the peak in the time-dependent photoluminescence spectra to a Gaussian function.

**Single-particle emission spectroscopy**: Solutions of QDs in hexanes were diluted to single-particle concentrations ~100 pM, deposited on a cleaned microscope slide, and quickly encapsulated with a coverslip. Single-particle emission spectra were recorded under a wide-field scheme, as described previously(*20, 21*). Briefly, the sample was mounted on a setup based on a Nikon Ti-E inverted fluorescence microscope and imaged in the wide field with an oil-immersion objective lens (CFI Plan Apochromat $\lambda$ 100×, NA 1.45) under 488 nm illumination. Emission was split into two channels for the concurrent recording of single-particle images and spectra, with the latter enabled by inserting a dispersive prism into the light path.

**Kinetic Monte Carlo simulation**: Simulations of incoherent exciton hopping trajectories were performed with discrete hops on a 5760 nm x 5760 nm 2D hexagonal lattice with periodic boundary conditions with lattice constant 9 nm. For each trajectory, site energies were randomly assigned in accordance with the experimentally measured inhomogeneous broadening width, and trajectories were initiated at a random site of the 2D lattice. Hopping rates between pairs of sites were then calculated from the site energies, the site-to-site separation, the intrinsic spectral width of the sites, and the intrinsic Stokes shift, in a dipole-dipole coupling model (see **Supplementary Materials** for details). Approximately 20,000 trajectories were averaged for each set of parameters and the resulting mean diffusivities were used to estimate the diffusion extent within TRUSTED experimental windows as well as within the exciton lifetime.

**Acknowledgments**

We acknowledge William Tisdale, Gregory Scholes, Dan Oron, and Milan Delor for helpful discussions. We acknowledge Dr. Karen Bustillo from the National Center of Electron Microscopy for her support on using the Titan X microscope.

**Funding:**
TRUSTED measurements were supported by the "Photonics at Thermodynamic Limits" Energy Frontiers Research Center of the U.S. Department of Energy, Office of Basic Energy Sciences under award no. DE-SC0019140 (RY, TDR, NSG). Time-resolved emission and single-particle spectroscopy were supported by the STROBE Center for Realtime Imaging, a National Science Foundation Science and Technology Center under grant DMR 1548924 (RY, TDR, AAC, HHP, KX, NSG). Data analysis and interpretation were supported by the Center for Computational Study of Excited State Phenomena in Energy Materials (C2SEPEM) under the U.S. Department of Energy, Office of Science, Basic Energy Sciences, Materials Sciences and Engineering Division under contract no. DE-AC02-05CH11231, as part of the Computational Materials Sciences Program (RY, TDR, NSG). Sample preparation and characterization were supported by the National Science Foundation, Division of Materials Research (DMR), under Award Number DMR-1808151 (RMB, JCO, APA) and by the U.S. Department of Energy, Office of Science, Office of Basic Energy Sciences, Materials Sciences and Engineering Division, under Contract No. DE-AC02-05-CH11231 within the Physical Chemistry of Inorganic Nanostructures Program (KC3103) (CY, APA). Electrodynamics simulations and analysis were supported by the U.S. National Science Foundation under Award CHE-1954393 (SK, DJM). R. Y. acknowledges a Dreyfus Environmental Postdoctoral Fellowship from the Camille and Henry Dreyfus Foundation. T. D. R. acknowledges a National Science Foundation Graduate Research Fellowship (DGE 1106400). T.D.R., R.M.B., and J.C.O. acknowledge Kavli Energy NanoSciences Graduate Fellowhips. N. S. G. acknowledges a David and Lucile Packard Fellowship, a Sloan Research Fellowship, and a Dreyfus Teacher-Scholar Award.

**Author Contributions:**
Conceptualization: RY, TDR, CY, JCO, APA, NSG
Methodology: RY, TDR, DJM, KX, APA, NSG
Investigation: RY, TDR, RMB, AAC, HHP, CY, JCO, SK
Visualization: RY, TDR, RMB, AAC, SK, DJM, KX, NSG
Supervision: DJM, KX, APA, NSG
Writing—original draft: RY, TDR, NSG
Writing—review & editing: RY, TDR, RMB, AAC, HHP, CY, JCO, SK, DJM KX, APA, NSG

**Competing Interests**: All authors declare they have no competing interests.

**Data and Materials Availability**: All data needed to evaluate the conclusions in the paper are present in the paper and/or the Supplementary Materials.

**Figures and Tables**

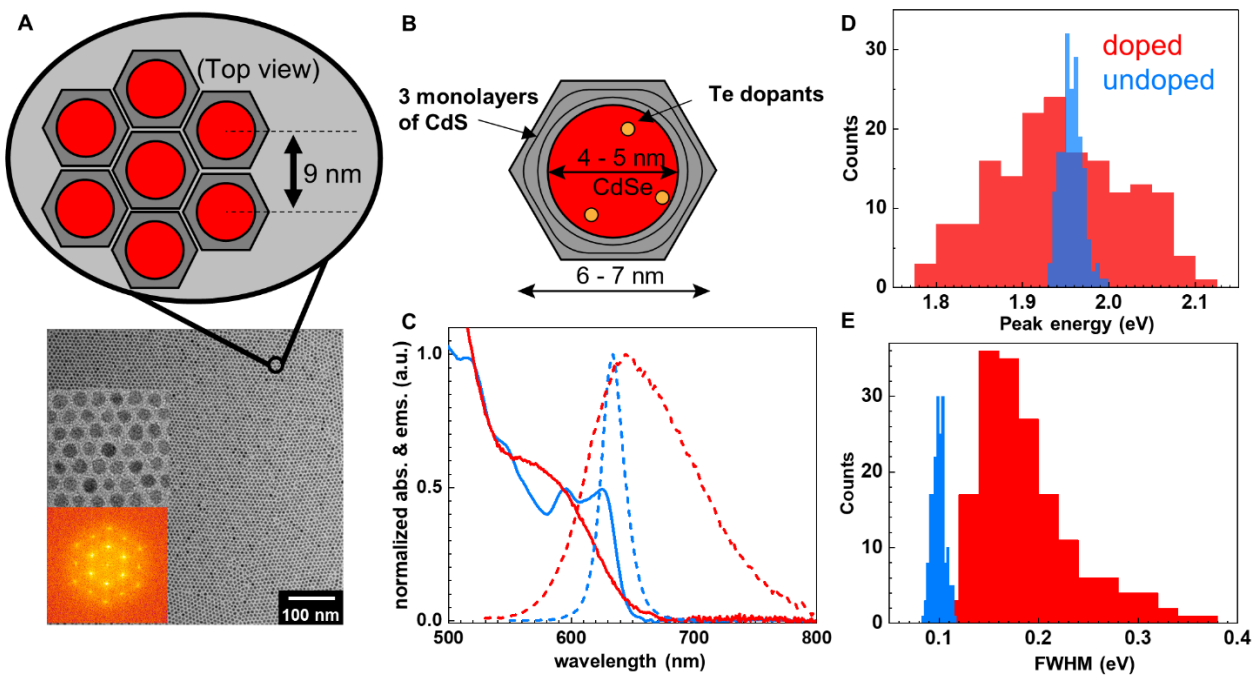

**Figure 1. Configuration and spectral properties of Te-doped CdSe/CdS QDSLs.** Te dopants in CdSe/CdS QDs broaden the absorption and emission spectra by increasing both the intrinsic linewidth and bandgap energy heterogeneity. (A) The hexagonal prism shape of the CdS shell promotes formation of a monolayer QD hexagonal SL with 9 nm lattice spacing, in which we investigate exciton migration dynamics, as shown in TEM and the schematic cartoon. Insets include a zoomed-in region of the TEM image and Fourier transform of the TEM image. (B) Schematic of the internal composition of the QDs used, including 4-5 nm spherical CdSe shell, ~3 monolayers of CdS shell in a short, hexagonal prism shape to form a 6-7 nm QD size, and presence of Te dopants. (C) Steady state QD absorption in colloidal suspension (solid curves) and QDSL emission spectra (dashed curves). Doped QD spectra are indicated in red; undoped QD spectra are shown in blue. Single-particle emission spectroscopy was used to obtain (D) the emission peak energy distribution of doped and undoped QDs and (E) the emission spectra FWHM of doped and undoped QDs.

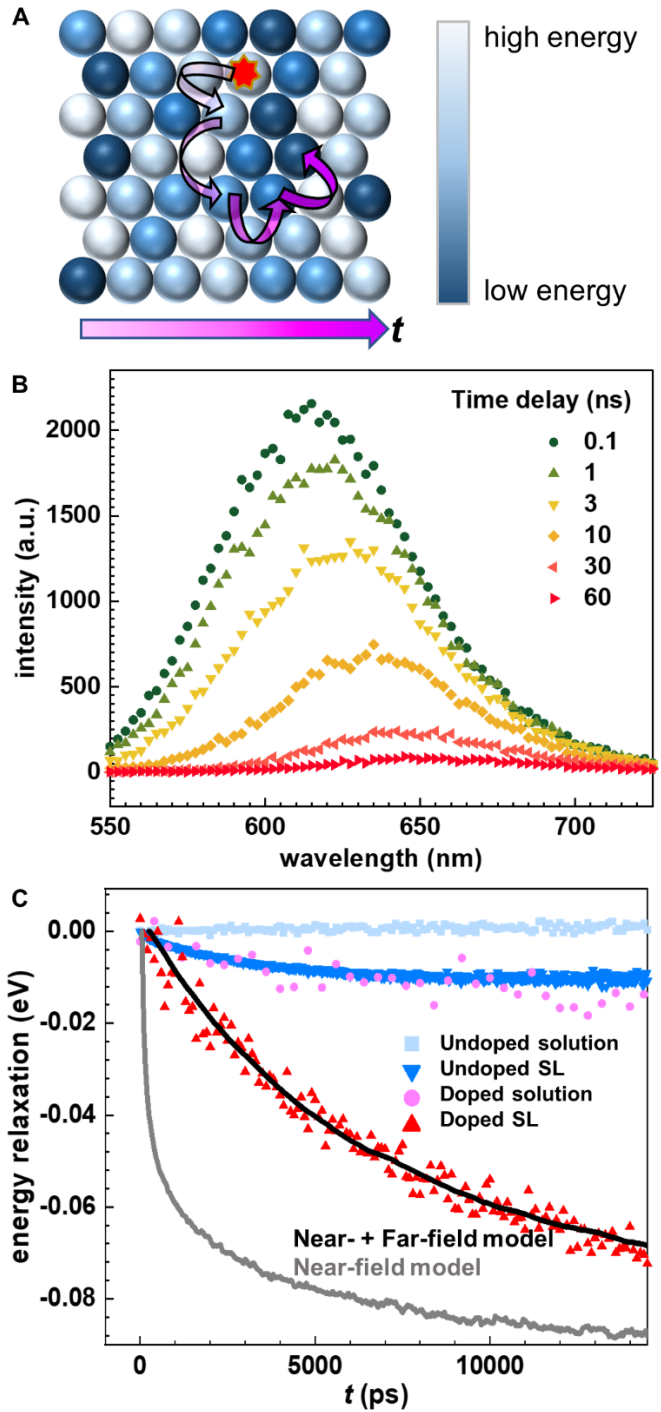

**Figure 2. Exciton migration dynamics tracked energetically by TRES.** (A) Cartoon of exciton (red star) hopping from a high-energy site (pale blue shades) to a low energy site (dark blue shades). Time is indicated with increasingly deep magenta. (B) Time-dependent photoluminescence spectra at select time delays after photoexcitation of a CdSe:Te/CdS QDSL. The emission peak energy shows an increasing redshift with increasing time delay. (C) Time-resolved emission spectra of doped QD solution (pink circles) and superlattices (red triangles) track the decay of mean exciton energy. Undoped solution (light blue squares) and superlattice (dark blue triangles) counterparts are also shown for reference. Simulated energy decay from spatial-spectral dynamics of near-field (gray and near-field plus far-field (black) models are also shown.

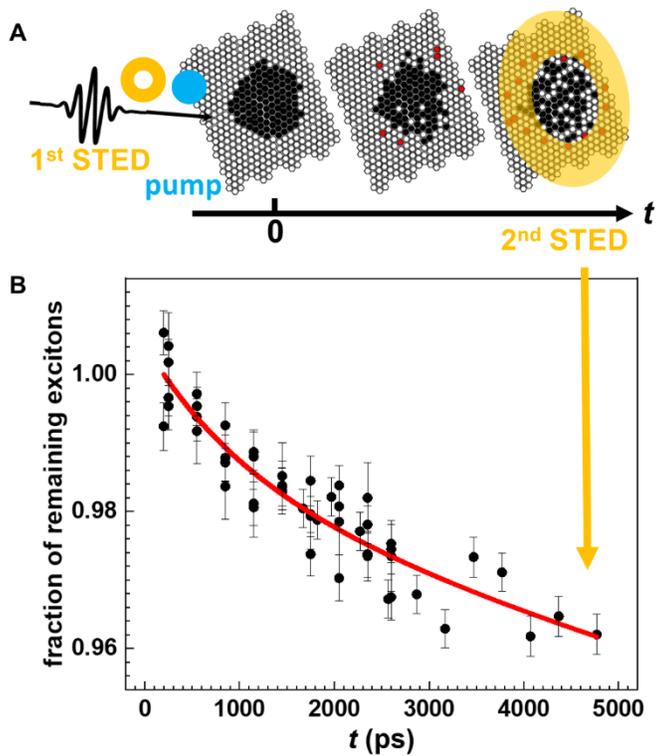

**Figure 3. Exciton migration dynamics tracked spatially via TRUSTED.** (A) Schematic diagram of a TRUSTED protocol. Black circles are excitons on the QDSL lattice (otherwise open circles) excited by the pump laser pulse and not quenched by the first STED laser pulse. Red circles correspond to excitons that migrate into the (yellow) annular area illuminated by the STED laser pulse intensity that are thus quenched upon the time-delayed arrival of a second STED pulse. (B) Measured TRUSTED fraction of remaining excitons (black points) decays as a function of time delay between first and second STED pulse arrivals are fit (red) using an exponentially decaying diffusivity model. Decreasing fraction corresponds to exciton diffusion.

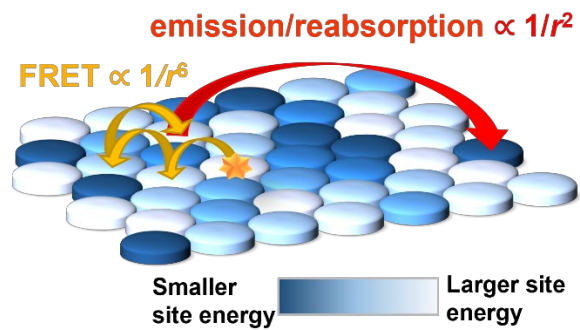

**Figure 4. Depiction of composite energy transport mechanism.** Schematic illustrating the inclusion of both near-field FRET hops and the occasional far-field emission/reabsorption over a heterogeneous energy landscape made up of sites with smaller (dark blue) and larger (white) exciton energies.

Supplementary Materials for

**A composite electrodynamic mechanism to reconcile spatiotemporally resolved exciton transport in quantum dot superlattices**

Rongfeng Yuan *et al.*

*Corresponding author. Email: nsginsberg@berkeley.edu

**This PDF file includes:**

Supplementary Text
Figs. S1 to S16
Tables S1 to S2
References (45 to 58)

1. **Sample preparation**:

   **Substrate.** Glass coverslips were sonicated for 10 minutes in successive baths of acetone, isopropyl alcohol, 2% Hellmanex solution in NANOpure deionized water, and 2X rinses in NANOpure deionized water, and immediately dried with filtered nitrogen flow. Coverslips were further cleaned with oxygen plasma in a reactive ion etch chamber.

   **Chemicals.** Cadmium oxide (CdO, ≥99.99%, Sigma Aldrich), octadecylphosphonic acid (ODPA, 99%, PCI Synthesis), trioctylphosphine oxide (TOPO, 99%, Sigma Aldrich), trioctylphosphine (TOP, 99%, Strem), selenium powder (Se, 99.999%, Sigma Aldrich), tellurium powder (Te, 99.8%, Sigma Aldrich), oleic acid (OA, 90% tech grade, Sigma Aldrich), oleylamine (OAm, 70% tech grade, Sigma Aldrich), 1-octadecene (ODE, 90% tech grade, Sigma Aldrich), n, n-dimethylformamide (DMF, anhydrous 99%, Sigma-Aldrich), toluene (anhydrous 99.5%, Sigma Aldrich), acetone (anhydrous, 99.5%, Sigma Aldrich), hexanes (mixture of isomers, anhydrous 95%, Sigma Aldrich), n-octane (98%, reagent grade, Sigma Aldrich). 0.2 M Cd(oleate)$_2$ in ODE precursor was prepared using a previously reported procedure(*46*). Briefly, appropriate amounts of CdO, OA, and ODE were degassed for about two hours under vacuum at 110 °C until all gases and water had evolved. The reaction was then heated to 240 °C under argon for ~ 30 minutes until a clear, slightly yellow, solution forms. The flask was then cooled to 110 °C and degassed a second time for an additional 1-2 hours to remove additional water. The solution was stored in an argon-filled glovebox.

   **Wurtzite CdSe:Te{5%} Nanocrystals.** Te-doped CdSe core nanocrystals were synthesized by modifying a similar procedure for undoped CdSe nanocrystals(*47*). Briefly, 120 mg CdO, 560 mg ODPA, 6 g TOPO were degassed under vacuum in a 50 mL round-bottom flask at 150 ºC for 1 hour. The mixture was then heated to 320 ºC under argon gas and maintained at 320 ºC until solution turned clear indicating formation of a complex. Upon completion of complex formation, 3 g of TOP was injected in the solution and then the temperature was raised to 360 ºC. A solution of the appropriate ratio of Se and Te totaling 1.5 mmol in 0.72 g TOP was swiftly injected when temperature approached 360 ºC. Upon injection, CdSe nanocrystals were formed and were allowed to grow for 90 seconds to produce large, uniform cores. Solution was then rapidly cooled with compressed air to 120 ºC before injecting 10 mL of anhydrous toluene and transferring to an inert atmosphere. The nanocrystals were cleaned in an inert environment with successive precipitation and redispersion using acetone and hexane respectively and were stored in 3ml of hexane. Concentration were determined by a previously reported empirical formula(*48*) and size was determined via TEM image analysis.

   **Hexagonal CdSe:Te{5%}/CdS core/shell Nanoplates:** A hexagonal CdS shell typically 3 monolayers thick was grown around the Te-doped CdSe nanocrystals(*49*). The anisotropic shell growth was grown by first degassing under vacuum 3 mL of ODE, 3 mL of OAm and 100 nmol of CdSe cores in a 25 mL round-bottom flask at 110 ºC for 30 mins. The solution was then heated to 310 ºC under Ar. When the temperature reached 240 ºC, injections of Cd and S precursor solutions in ODE started at a rate of 3 ml/hr using a syringe pump. The Cd precursor solution contained ~4 mL of 0.20 mM Cd(OA)$_2$ and the S precursor contained ~25 mg of S powder in ~3 g of TOP. Upon completion of the precursor injections, the reaction was maintained at 310 ºC for 10 minutes, then cooled rapidly using compressed air to room temperature. Successive precipitation and redispersed using acetone and hexane respectively were done to clean the nanocrystals capped with OA and OAm before they were stored in 2 mL hexane in an inert atmosphere.

QDs without Te dopants were also prepared. To match the peak emission wavelength of the Te-doped QDs, we increased the undoped CdSe core size and kept the same 3 monolayer CdS shell.

**Self-assembly of Nanocrystal Superlattice**: Formation of nanocrystal superlattice was performed using a previously published self-assembly method(*50*). Monolayer nanocrystal superlattices were formed at the liquid-air interface at room temperature. The nanocrystal stock solution was diluted in octane to a proper concentration to obtain monolayer coverage. Approximately, 50 uL of diluted nanocrystal solution was drop-casted on top of 1mL of DMF in a $1cm^2$ square Teflon well. The well was covered with a glass cover slip to slow solvent evaporation rate overnight. The self-assembled superlattice was scooped onto a plasma cleaned glass cover slip for optical measurements or a Cu 400 mesh standard carbon TEM grid for structural characterization. For optical measurements, the substrate is glass coverslips.

2. **Sample characterization**

<u>**Single QD photoluminescence spectra.**</u> Steady state photoluminescence spectra of 170 individual doped QDs and 182 individual undoped QDs were recorded according to the procedure described in the Materials and Methods section. Below, three examples of 5% Te doped individual QD photoluminescence spectra (orange, blue, and red curves) and the average of 170 such spectra (dashed red curve) are plotted. The statistics of the peak energy and FWHM distributions in Figure 1d and e are extracted from these 170 individual PL spectra.

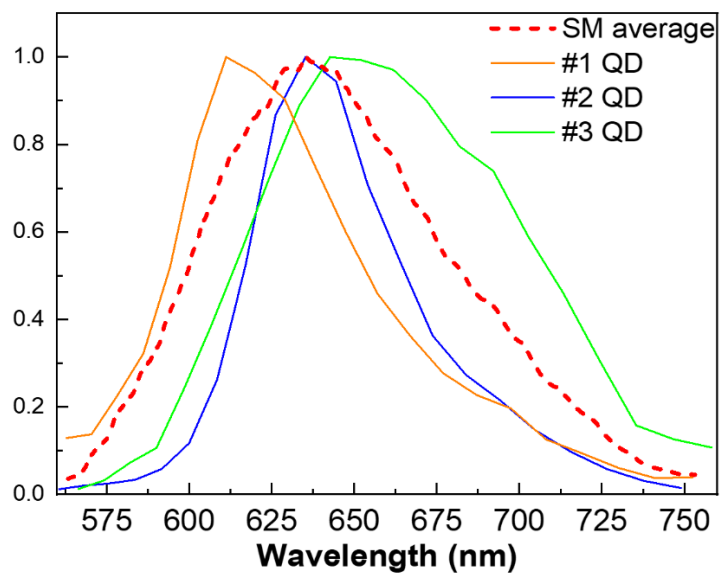

**Figure S1. Example of three individual 5% Te doped QDs PL spectra and the average of 170 such spectra.**

**Absorption and emission spectra of undoped and 5% Te doped QD.** Steady state absorption and emission spectra of sample in hexane solution verify the successful incorporation of the Te dopants within the CdSe matrix, shown in **Figure S2**. Absorption spectra were collected with a Shimadzu UV-3600 double beam spectrometer operating with 1 mm slits. Photoluminescence (PL) spectra were obtained with a 470nm excitation laser and Picoquant FluoTime 300 Fluorometer. The absorption spectra of undoped CdSe/CdS and doped CdSe:Te/CdS systems are shown as solid curves and their PL emission are shown as dashed curves. The absorption spectra are scaled to aid in comparison, and the emission spectra are normalized to their peak emission intensity.

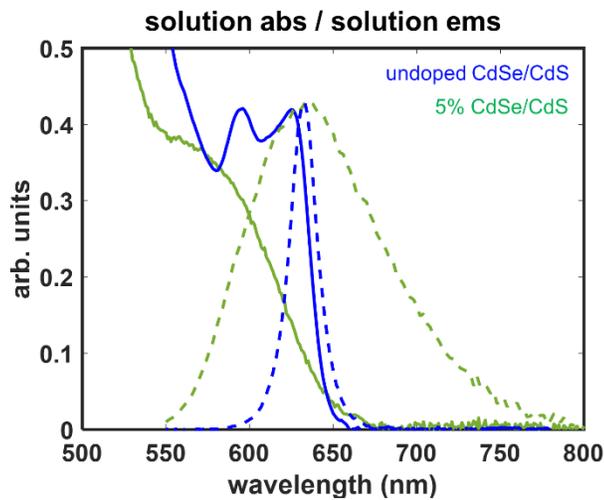

**Figure S2**: **Steady state absorption and emission spectra of just undoped and 5% doped solution.**

Both the undoped QD solution's and the 5% Te doped QD solution's PL emission peaks at ~632 nm. The width of the doped emission spectra is substantially wider. The FWHM of the undoped CdSe/CdS QDs in solution is ~65 meV, and for the doped profiles, ~286 meV.

When measured in the superlattice solid-state, the PL emission of both systems redshift (shown in **Fig. 1** in the main text). The undoped QDSL PL emission peaks at ~634 nm, and the doped emission profiles peak at 646 nm. The FWHM of the undoped QDSL emission profile is ~72 meV, and for the doped profiles, ~284 meV.

For the 5% CdSe:Te/CdS QD solution, the absorption onset is smeared and renders assigning a first exciton peak less obvious. Here we used a crude way to estimate the Stokes shift. The absorption lineshape was fit with a double Gaussian function, one to capture the shoulder (sloping baseline) and one to capture the absorption onset. This yields a 586 nm absorption peak and a 160 meV Stokes shift, which is in line with a previous estimate(*24*).

**X-ray Diffraction:** X-ray diffraction was measured of nanocrystal sample drop-casted on a silicon substrate. XRD was collected on a Bruker Phaser D2 diffractometer with a Cu kα source operated at 30kV and 10 mA with a 160 SSD detector. Diffraction measurement was done from 20° to 60° 2Θ with increments of 0.01° with an integration time of 3 s/step.

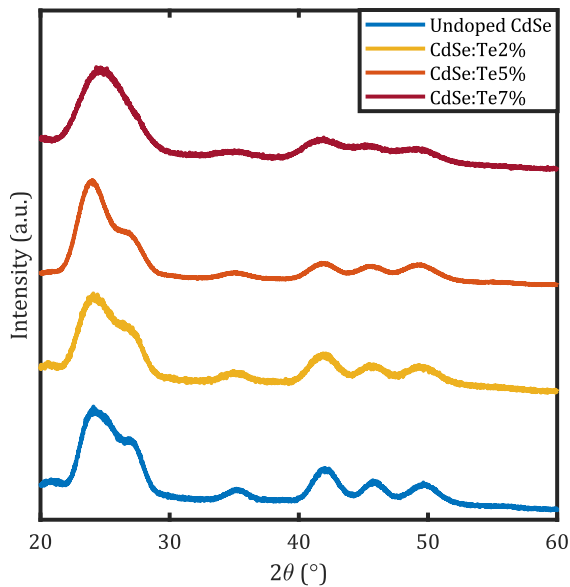

**Figure S3**: **X-ray diffraction pattern of Te-doped CdSe/CdS core shell quantum dots.**

From the powder X-ray diffraction patterns, we observe that the incorporation of Te atoms in the CdSe matrix does not substantially affect the CdSe crystal structure of the nanocrystal. CdSe and CdTe quantum dots can be synthesized to have either a wurtzite(wz-) or a zinc blende(zb-) crystal structure. The quantum dots used in the work have been prepared by modifying a known high-temperature synthesis of wz-CdSe(*47*). It has been reported that when synthesized at high temperatures, the zb-CdTe crystal structure is favored over wz- CdTe(*51*). All wurtzite peaks are present for the Te-doped and undoped CdSe quantum dots. As we increase the amount of Te dopant in the CdSe host matrix, we observe a broadening of the diffraction peaks due to minor change in the lattice parameter due to Te having a larger atomic radius than Se. For the CdSe:Te{7%} sample which has the most Te dopants, the broadening of the peaks lead to a merging of the 101 and 002 peaks, the presence of the other wz-CdSe peaks (102,110,103,112) show that the Te dopant at these dopant levels does not substantially change the crystal structure of the wz-CdSe nanocrystals. We also do not observe any signature zb-CdTe XRD peaks in the XRD measurements. Based on these measurements, we can conclude that the Te is likely incorporated into the CdSe matrix as a dopant as opposed to the formation of a CdSe:CdTe alloy.

**STEM-EDX Characterization and Analysis**: To investigate whether the synthesis strategy yielded the expected Te doping percentage in QDs, we conducted scanning transmission electron microscopy – energy dispersive X-ray (STEM-EDX) experiments. Here, each element (Cd, S, Se, Te) can be quantified by its corresponding X-ray spectral features. Within experimental error, STEM-EDX confirmed that the Te dopant percentage, (6.2 ± 2.3) % is the same as our expectation of 5% from the QD synthesis procedure.

Elemental composition analysis of nanocrystals was performed at the National Center for Electron Microscopy, Molecular Foundry, Lawrence Berkeley National Laboratory. Samples were drop-casted on ultrathin 400 mesh carbon Cu TEM grids and left under vacuum to remove any excess solvent and oxygen plasma treated. High-angle annular dark field HAADF-STEM and STEM-EDS maps was collected on the FEI TitanX 60-300 microscope using a FEI low-background double-tilt holder. HAADF-STEM was performed at 300 kV with a beam convergence semiangle of 10 mrad using a Gatan HAADF detector. STEM-EDS mapping was performed at 300 kV with a screen current of ~2 nA using four windowless silicon drift detectors with a total solid angle of 0.7 steradians and 140 eV energy resolution. STEM-EDS maps were collected using the Bruker Esprit software for ~5-10 minutes utilizing drift correction. Quantification of the elemental composition of each sample was done using the Bruker Esprit software using the Cliff–Lorimer method for each element: Cd K-series, Se K-series, S K-series, Te L-series.

**TEM Characterization and Image Analysis for Verification of superlattice formation**: Transmission Electron Microscopy (TEM) and High Resolution TEM (HRTEM) were used to characterize nanocrystal superlattices using a FEI Tecnai T20 S-TWIN TEM operating at 200 kV with a $LaB_6$ filament. TEM images were collected using a Gatan Rio 16IS camera with full 4k by 4k resolution using the drift correction feature. Sizing distribution curves were generated from TEM images of > 4000 nanoparticles using a custom written MATLAB script in which the details have been previously reported(*52*). Image analysis and Fourier transform of image was performed on ImageJ software.

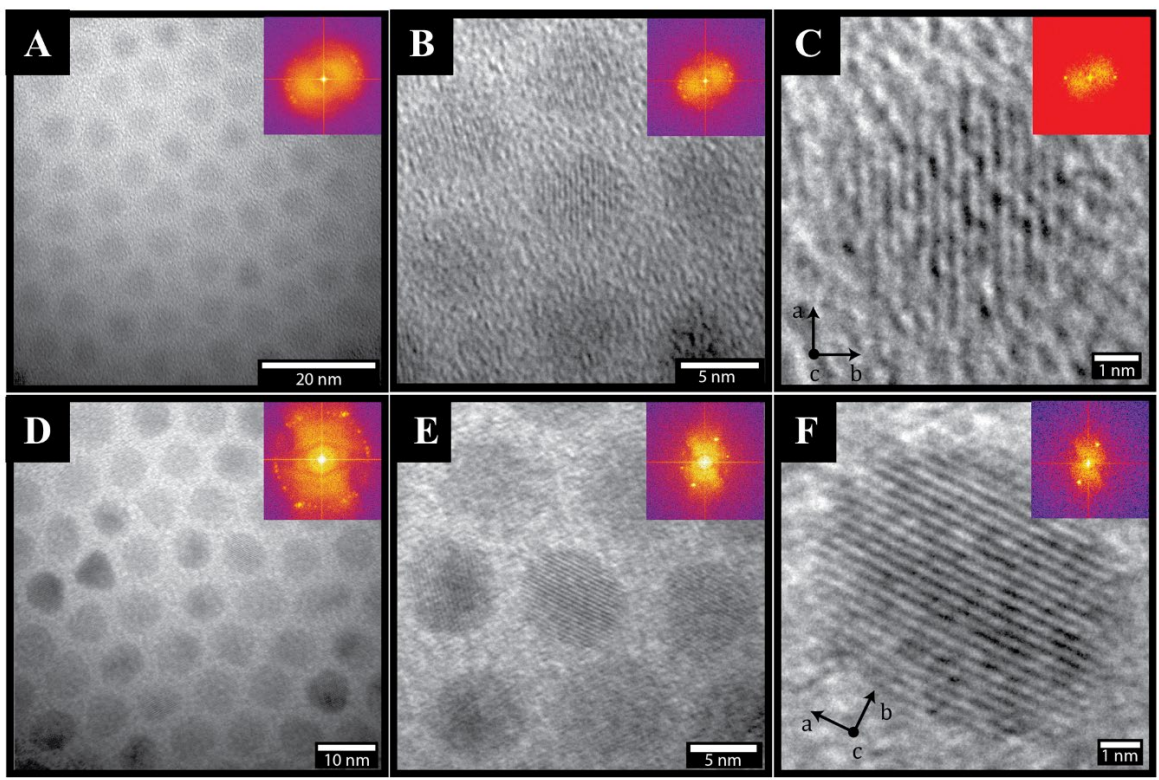

**Figure S4. HRTEM was performed on the QD superlattices to confirm the nanocrystal shape, ordering and crystallographic alignment.** For the CdSe:Te/3MLCdS samples studied in the main text we measured these highly spatially ordered superlattices (A). Based on the fast Fourier transform (FFT) inset of the image we also observe crystallographic ordering since there are discrete points in the FFT instead of rings that are often observed in disordered systems. By zooming into a smaller section of the image (B), we better observe the crystallographic alignment of these QDs as the discrete points are localized in a specific region showing that there is interparticle crystallographic alignment. The image of one CdSe:Te/3MLCdS QD (C) shows only two points in the FFT that match the regions in the FFT of the wider view images. From these images we observe that the QDs have both good spatial ordering and crystallographic ordering with the *c*-axis of the QD being perpendicular to the substrate and the a-axis mutually aligned in-plane. We performed these same analysis for the CdSe:Te/6MLCdS QDs and observed a similar trend (Fig D-F). Since the CdSe: Te/6MLCdS QDs are larger than the CdSe:Te/3MLCdS, the HRTEM images have better resolution and contrast, making it easier to see the QD edges and understand the nanocrystal shape. From the HRTEM of the CdSe:Te/6MLCdS sample (Fig D-F), we can see that these QDs have a hexagonal prism shape as discussed in the main text as opposed to an isotropic circular shape.

**Tapping mode AFM imaging for Verification of superlattice formation**: AFM images were acquired using an Asylum MFP-3D with TAP 150 AL silicon AFM probes from Ted Pella, Inc. Scans were performed with 31.0 nm per pixel resolution at a 0.5 Hz scan rate with 0.73 V driving amplitude.

AFM height mapping is useful in determining whether the large area superlattice structures we investigate are monolayers, bilayers, or higher order thicknesses. **Figure S5** is a representative AFM height map of a piece of superlattice film. The central, relatively featureless surface is approximately 6-7 nm in height and agrees well with the estimated 5 nm height of our CdSe:Te/CdS QDs (excluding the ligand layer). The additional lighter contrast features adorning the periphery of the superlattice fragment represent smaller subregions of bilayers, with an approximate height of 12-13 nm. The lightest contrast features are larger scale structures, possibly thicker disordered aggregates of QD material.

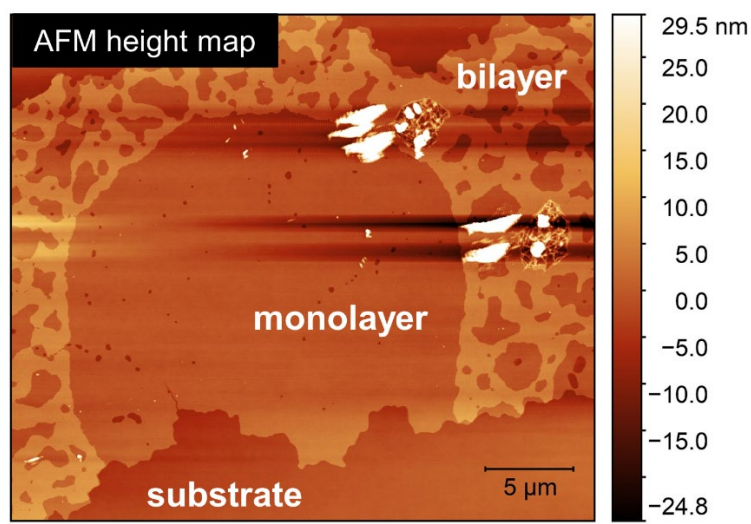

**Figure S5**: **AFM height image of a CdSe:Te/CdS superlattice fragment.** Labeled in the image are regions of monolayer and bilayer superlattice formation.

**Brightfield optical image of QDSL monolayer and bilayer**

Brightfield images were obtained by illuminating the sample with a while light LED and collecting the transmission with a ThorLabs CMOS camera. **Figure S6** shows an example image of a region of predominantly CdSe:Te/CdS monolayer superlattice (left) and a region of predominantly CdSe:Te/CdS superlattice bilayers (right). In both cases, owing to the 10s of nm thickness, we slightly defocus the image in order to more easily see the superlattice fragments. For TRUSTED experiments, we only examined monolayer locations in the whole sample.

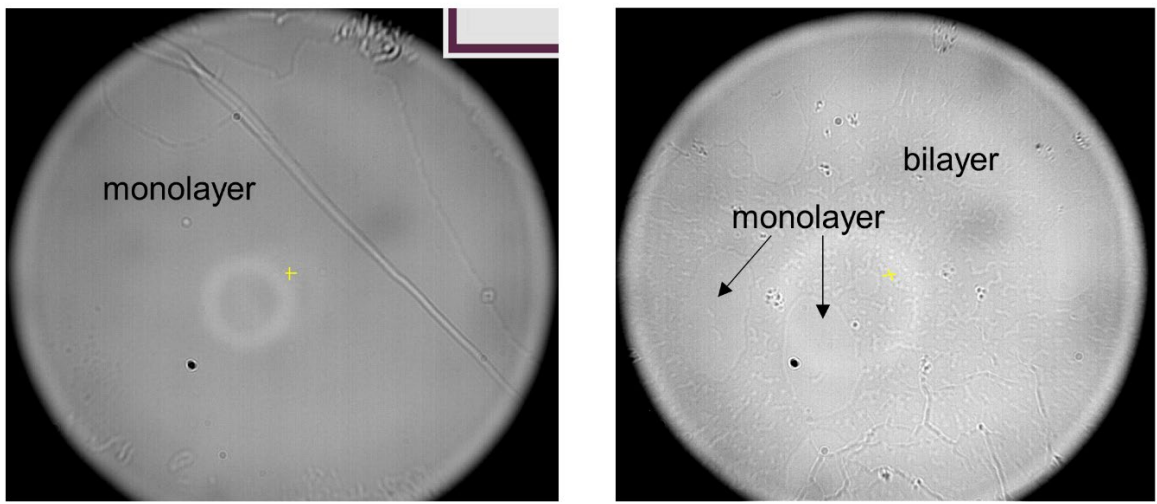

**Figure S6**: **Brightfield images of CdSe:Te/CdS fragments.**

## 3. TRES data processing and fitting procedure

**Figure S7** shows the time-dependent normalized PL spectra of the doped QDSL. The factory-set photomultiplier tube (PMT) calibration curve is also plotted in forest green. The PMT has a lower sensitivity to redder photons, but the signal-to-noise ratio of TRES is large enough that we did not observe a sloping baseline in the time-dependent PL spectra. Thus, it is unlikely that the correction factor would create an artificial redshift in TRES.

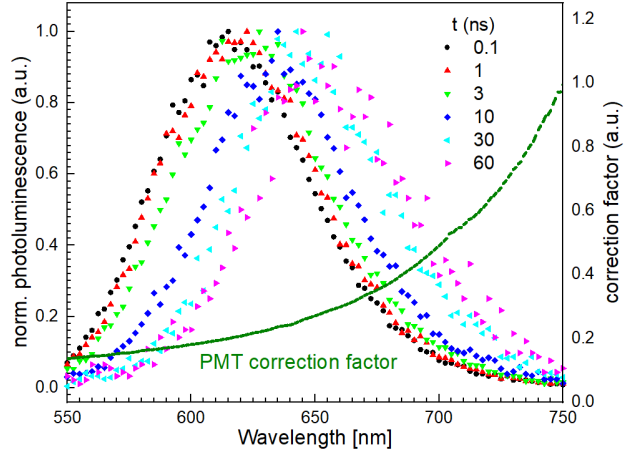

**Figure S7. The time-dependent PL spectra of the doped QDSL normalized to the same peak height at the same time delays.** The forest green dashed curve corresponds to the factory-set calibration curve of the PMT response.

According to Bässler(29), if we assume the density of site energies follows a Gaussian distribution of states, then from time-resolved emission traces one can estimate the inhomogeneous broadening components from the following expression, provided the energy relaxation in the system follows Boltzmann statistics and is within a "hopping" regime:

$$\Delta E = \Delta E_\infty (1 - e^{-k_{\Delta E} t})$$

*Eq. S1*

where $k_{\Delta E}$ is the rate constant corresponding to the rate of energy change in the system and $\Delta E_\infty$ represents the equilibration energy, which can be related to the inhomogeneous broadening via:

$$\Delta E_\infty = \frac{-\sigma_{ih}^2}{k_B T}$$

*Eq. S2*

where $\sigma_{ih}$ is the inhomogeneous broadening component of the spectral linewidth. Bässler demonstrates that **Equation S2** is a result of assuming an excitation reaches an "equilibrium" amongst lattice sites within its lifetime. The overall decrease in energy as reported by TRES is ~ 120 meV; thus, $\sigma_{ih}$ is 56 meV. Note that the above treatment is only strictly valid if we assume the exciton transfer rate has an energy dependence that follows a Boltzmann distribution. This is unlikely the case for the QDSL in the strict sense. **Figure S8** shows the single exponential fit of the time-dependent energy loss due to exciton transport. The fact that the fit is not good hints the limitation on this model, but this model still provides a sense of the relative disorder in the energetic landscape.

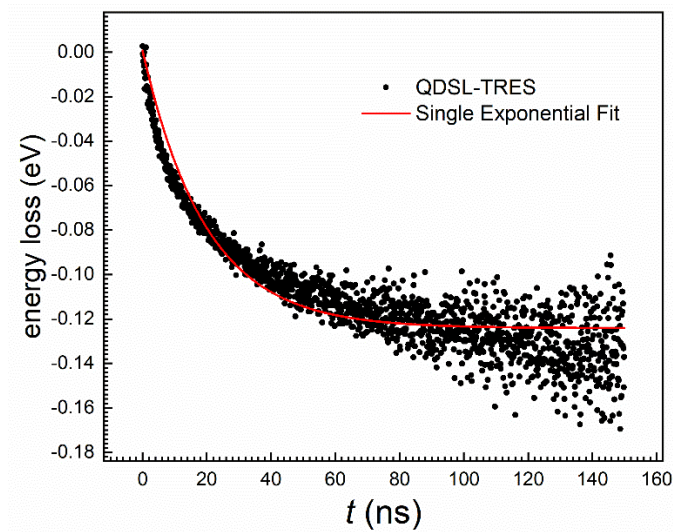

**Figure S8. Decay of mean exciton energy of a Te doped CdSe/CdS QDSL measured by time-resolved emission spectroscopy (TRES).**

## 4. Spatiotemporal measurement of exciton transport

### STED resolution determination

One large source of uncertainty in assigning a diffusivity from TRUSTED datasets, here, is our measurement of the initial FWHM of the exciton population resulting after the action of the pump and first STED pulse. According to our resolution determination measurement, the FWHM of our initial exciton distribution is ~220 nm, where the fitting model uncertainty (weighted by the uncertainty in the data) bounds this estimate between 215 and 230 nm. See **Table S1** for diffusivity estimates using these initial FWHM estimates.

The TRUSTED fitting scheme(*22*) used to extract a diffusivity requires as an input parameter the FWHM of the exciton distribution after the action of STED1. We determined this value by imaging a sub-diffraction limited sized fragment of a 5% CdSe:Te/CdS superlattice at a series of different STED powers. **Figure S9** presents a plot of the FWHM vs STED power (measured at the sample plane), where the FWHM values are the result of fitting a line-cut of the imaged fragment using a Gaussian. The STED power we employ in our TRUSTED measurements is 25 µW at the sample plane (at 200 kHz repetition rate and ~120 ps pulse duration). We fit the data using **Equation S3** below (red curve) and generate a $2\sigma$ error in the fit by considering both the variance in the fitting as well as the uncertainty in the data, which is the uncertainty of the linecut fits used to extract the FWHM from the imaged QDSL fragment (red shaded region).

$$FWHM_{eff} \approx \frac{FWHM_c}{\sqrt{1 + \frac{I_{STED}}{I_{sat}}}}$$

*Eq. S3*

According to **Figure S9**, at a 25 µW STED power we assume our resolution to be 220 nm, with an upper and lower bound of 230 and 215 nm, respectively. Fitting the TRUSTED datasets with these upper and lower bounds of the FWHM estimate, with both a constant diffusivity and time-dependent diffusivity fit, yields the results presented in **Table S1**. **Table S1** demonstrates that in all cases, the diffusivity extracted from our TRUSTED measurement is on the order of $10^{-3}$ cm$^2$/s. With regards to the time-dependent diffusivity model, we note that the time-constant of the decay is nearly identical across our three initial FWHM guesses, and the ratio of $(D_o + D_c)$ to $D_c$ ranges between 3.3 and 3.9. As $D_c$ is the hypothetical diffusivity at equilibrium (i.e., as time approaches infinity), the diffusivity can conceivably decay by a factor between 3.3 and 3.9, provided an exciton lives long enough to reach equilibrium.

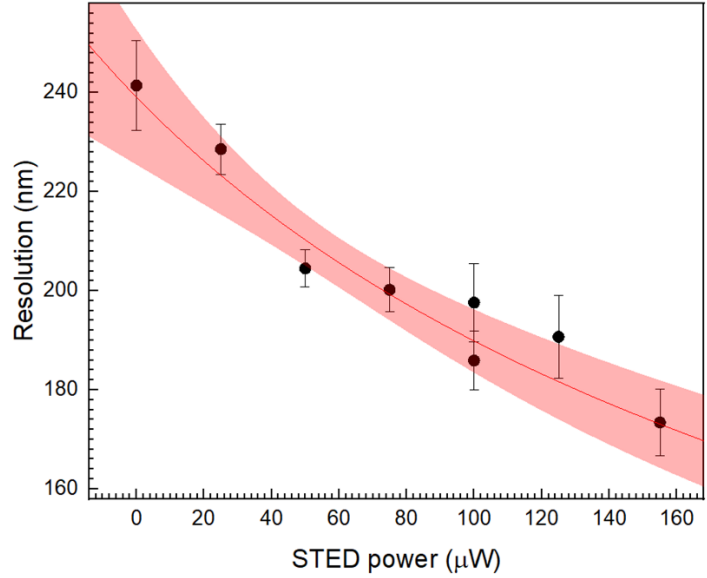

**Figure S9**: **Resolution vs STED power of an isolated 5% CdSe:Te/CdS QDSL fragment.** The red curve is the result of a fit to **Eq S3**, and the pink shaded region represents the 2σ range of the error of the fit weighted by the uncertainty in the data.

| FWHM(nm) | $D_o$ (x10$^{-3}$ cm$^2$/s) | $D_c$ (x10$^{-3}$ cm$^2$/s) | $k_d$ (ps$^{-1}$) | model |
|---|---|---|---|---|
| 215 | 4.0 ± 2.4 | 1.4 ± 1.0 | 0.0008 ± 0.0010 | $D_o e^{-k_d t} + D_c$ |
| 220 | 4.4 ± 2.7 | 1.7 ± 1.2 | 0.0008 ± 0.0011 | $D_o e^{-k_d t} + D_c$ |
| 230 | 5.5 ± 3.8 | 2.4 ± 1.5 | 0.0009 ± 0.0012 | $D_o e^{-k_d t} + D_c$ |
| 215 | 2.3 ± 0.2 | - | - | $D_o$ |
| 220 | 2.7 ± 0.3 | - | - | $D_o$ |
| 230 | 3.7 ± 0.3 | - | - | $D_o$ |

**Table S1. TRUSTED fitting results with upper and lower uncertainty of initial FWHM.**

## TRUSTED control experiments

To substantiate our confidence in the TRUSTED measurement, a few key control measurements were required to rule out potential non-diffusion related contributions to the TRUSTED observable.

### Exciton-exciton annihilation or Auger-Meitner recombination

Exciton density-dependent contributions to the non-radiative decay of the exciton population, such as exciton-exciton annihilation or Auger-Meitner recombination, could be mistaken for migration due to relative changes in the spatial profile leading to more pronounced overlap with the STED pulse. Provided the timescales associated with such a change to the exciton profile are within our observation window (e.g., 200 to 4800 ps) this could lead to a time-dependent change in the quenching action of the second STED pulse. To address this, we collected the PL from 25 different spatial locations on a 5% CdSe:Te/CdS monolayer at several pump powers. We plot the average of the collected PL counts from the different spatial locations as a function of pump power measured at the sample in **Figure S10**. We fit a linear trend to the lowest 5 powers and extrapolate across the entire range of pump powers measured. At powers higher than 5 nW, we note that the average PL counts begin to deviate substantially from the expected trend, which is a characteristic sign of Auger recombination or exciton-exciton annihilation.

TRUSTED measurements were conducted at 2 nW and 5 nW powers, which correspond to excitation densities of $\sim3.1\times10^{-5}$ excitations/nm$^3$ and $\sim7.7\times10^{-5}$ excitations/nm$^3$, respectively, indicated as arrows in **Figure S10.** The estimated diffusivities are $(2.5 \pm 0.7) \times 10^{-3}$ cm$^2$/s and $(1.7 \pm 0.3) \times 10^{-3}$ cm$^2$/s, respectively, and they are within error of one another, suggesting we are operating within a linear excitation regime.

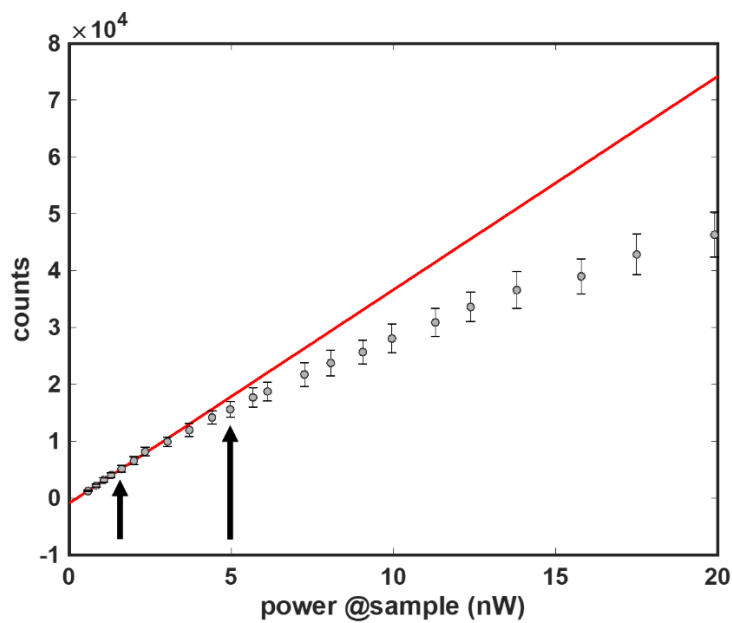

**Figure S10. PL counts vs. pump power measured at the sample plane.** The red curve is a linear fit determine by the lowest 5 power data points. The black arrows indicate 2 and 5 nW excitation, which are the two powers used in TRUSTED.

## Exciton transport in 2% and 7% Te doped CdSe/CdS core-shell quantum dot superlattices

We characterized the exciton transport properties of QDSL monolayers of 2% and 7% Te doping levels as well. These two QDs share the same CdSe core size and CdS shell thickness as the 5% Te doped CdSe:Te/CdS QDs. We do want to note that the 2% and 7% samples are from a different batch of nanocrystals, precluding making quantitative comparisons between their time-resolved emission traces as well as TRUSTED datasets and those of the 5% samples.

First, we measured the QD solution absorption as well as QD solution and superlattice emission spectra of the 2% and 7% Te doped QD show in **Figure S11** below. For reference, they were plotted together with the undoped and 5% Te doped QD sample presented in the main text.

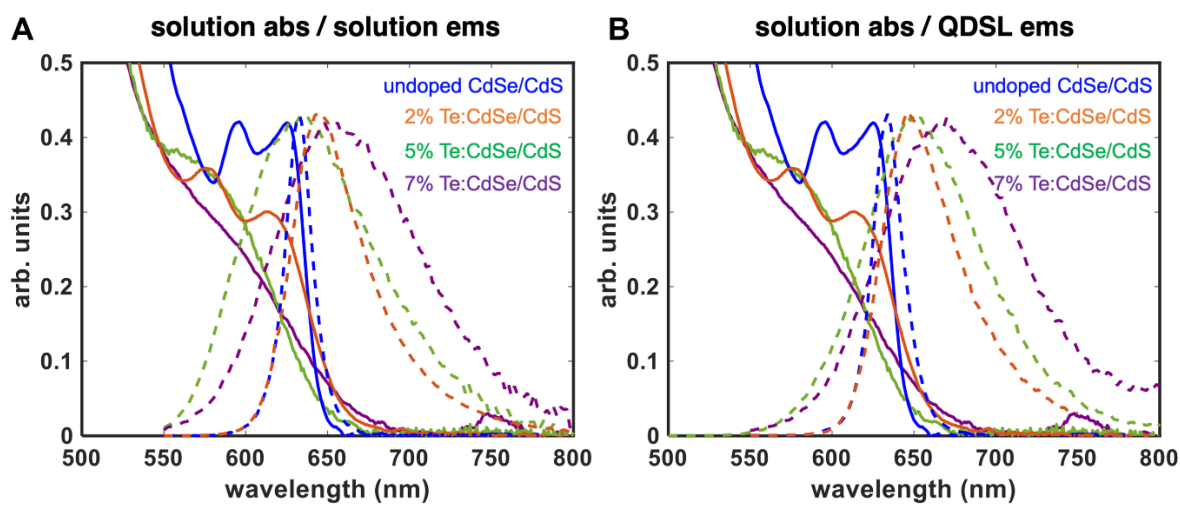

**Figure S11: Steady state absorption and emission spectra of undoped and 2, 5, and 7 % Te-doped CdSe/CdS QDs.** (A) Absorption and emission spectra of all 4 systems, collected in hexanes solution. (B) The same solution absorption spectra as (A), but the emission spectra are collected from the superlattices of all 4 systems.

Using TRES (**Figure S12**), we tracked the decay of mean exciton energy and found that similar to 5% Te doping, a larger dynamic redshift is present when the QDs are arranged in superlattice compared to the QDs suspended in solution. Like **Figure 2** in the main text, this suggests that there is energy transport in the QDSLs made from 2% and 7% Te doped CdSe/CdS QDs.

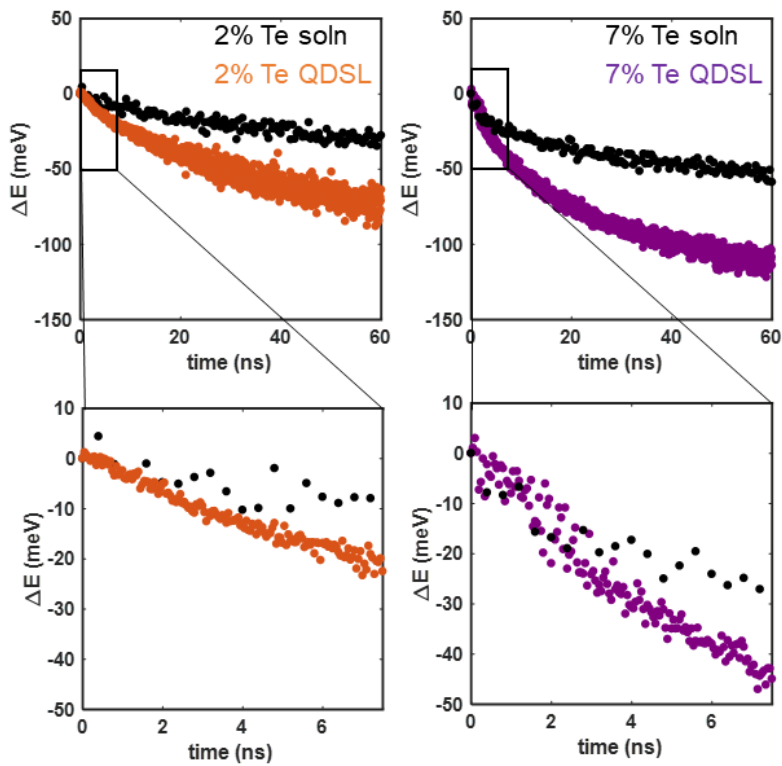

**Figure S12**: **TRES of 2, 7% Te doping of CdSe/CdS QDs in solution (black data) and in superlattice (colored data).**

Finally, using TRUSTED, we measured exciton diffusivities in the 2 and 7% Te-doped QDSL monolayers. The 2% and 7% data are presented in **Figure S13**. The TRUSTED data for each sample are shown, assuming a constant diffusivity model, and we obtain diffusivities of $(1.3\pm0.2)\times10^{-3}$ cm$^2$/s and $(1.3\pm0.5)\times10^{-3}$ cm$^2$/s for the 2% and 7% doped samples, respectively. We did not attempt a fit to the data with a time-dependent diffusivity model given the relatively small change in the normalized detection volume fluorescence. The relatively larger error in the 7% data is a consequence of the lower PL quantum yield of the 7% QDSL monolayers. As mentioned above, we refrain from making a direct comparison between the 5% dataset and the 2% and 7% data given we suspect batch-to-batch variability in the nanocrystal preparation.

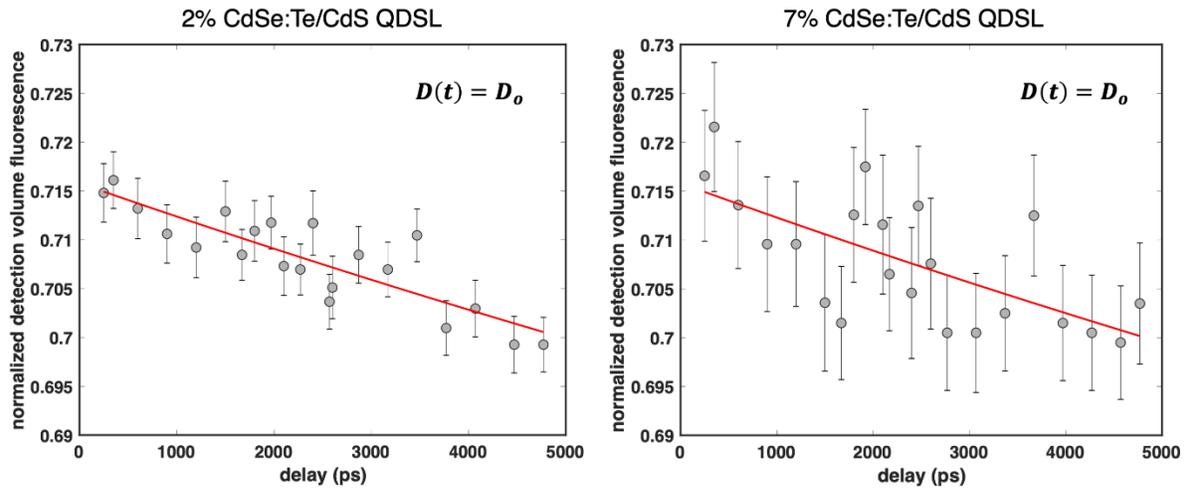

**Figure S13**: **TRUSTED measurements of CdSe:Te/CdS superlattice monolayers for the 2% (left) and 7% (right) Te-doping.**

One might anticipate the TRUSTED data to follow a trend as a function of the Te doping level, as is seen for the steady-state spectra and TRES measurements. It is important to note, however, that an increasing magnitude of energy relaxation in TRES does not necessarily imply faster or slower exciton transport. In fact, the energy relaxation rate is similar among the 2%, 5% and 7% doped CdSe:Te/CdS QDSL samples, especially when compared to the relaxation rate of the undoped sample as is shown in **Figure S14**.

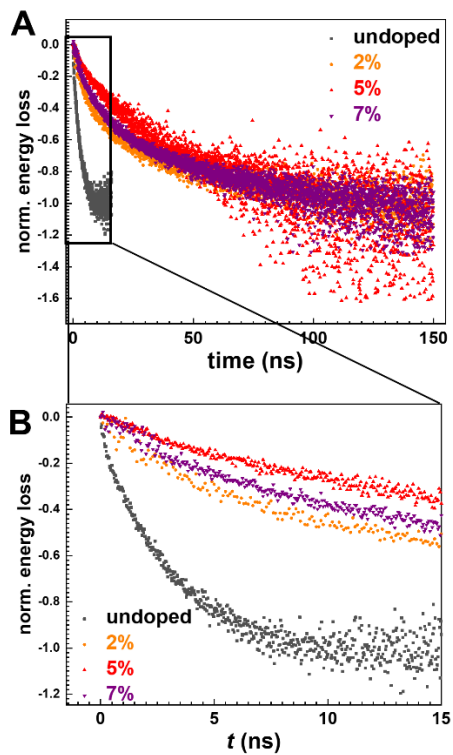

**Figure S14**. **Time-resolved emission of variously doped QDSL samples.** (A) Time-resolved emission data for the undoped (black), 2% doped (orange), 5% doped (red), and 7% doped (purple) QDSL samples are normalized to [0, -1]. Data extending beyond 20 ns for the undoped QDSL sample are omitted, as the lifetime of the undoped QDSL is substantially shorter than the doped QDSLs. (B) The same data are presented as in (A) but with the range limited to 15 ns.

## 5. FRET model for determining exciton diffusivities

Revisiting the equations describing the rate of energy transfer in a FRET framework, the rate constant is:

$$k_{FRET} = \frac{1}{\tau_D}\left(\frac{R_o}{r}\right)^6$$

*Eq. S4*

where $\tau_D$ is the radiative lifetime of the donor chromophore, $r$ is the distance between the centers of the point transition dipoles used to approximate the spatial extent of the electronic transition densities corresponding to the ground-to-excited state transitions, and $R_o$ is the "FRET radius", which is the distance at which the energy transfer rate is 50% efficient. $R_o$ packages up many terms, and is defined as:

$$R_o^6 = \frac{9\log(10)\,\Phi_D \kappa^2}{128\pi^5 \eta^4}\int \sigma_A(\lambda)\overline{F_D(\lambda)}\lambda^4 d\lambda$$

*Eq. S5*

where $\Phi_D$ is the fluorescence (PL) quantum yield of the donor chromophore, $\eta$ is the index of refraction of the medium, $\kappa^2$ is the dipole-dipole orientation factor, $\sigma_A(\lambda)$ is the wavelength-dependent absorptivity of the acceptor chromophore species, and $F_D$ is the fluorescence emission spectrum normalized by its area. Note that $\int \sigma_A(\lambda)\overline{F_D(\lambda)}\lambda^4 d\lambda$ is referred to as the spectral overlap integral and often denoted as *J*.

For our CdSe:Te/CdS system, we establish estimates of the various parameters used to determine $R_o$. We experimentally measured $\Phi_D = 23\%$ using photothermal deflection, as in Supporting Ref (*46*). We assume the index of refraction $\eta$ to calculate FRET-based transport to be between 1.8 - 2.0 based on Dement *et. al.*'s investigation of CdSe/CdS QD thin films(*53*).

The last two parameters, the spectral overlap integral and $\kappa^2$, are potentially the greatest sources of uncertainty. For $\kappa^2$, a conservative estimate is typically to assume that the transition dipole moments of all chromophores in the ensemble are isotropically distributed, such that $\langle\kappa^2\rangle = 2/3$. In a highly ordered superlattice, however, the orientation of QD transition dipole moments could be more aligned, yielding more optimal configurations than 2/3 for $\langle\kappa^2\rangle$. Here, for an estimate of $R_o$, we use $\langle\kappa^2\rangle = 2/3$. To measure the extinction coefficient of QDs, we measured the absorption spectrum of a stock solution of CdSe:Te/CdS QDs with a 45 mg/mL concentration that is diluted 100-fold. Based on Striolo *et. al.*'s measurements using membrane osmometry as a benchmark for correlating CdSe nanocrystal diameter with molecular weight(*54*), and taking into account the relative amounts of Cd, Se, S, and Te in our system, we crudely estimate the molecular weight of our CdSe:Te/CdS nanocrystals to be 1000 kg/mol. Based on these values and **Eq. S5**, the theoretical $R_o$ is ~ 3.7 nm. This value can be higher with a more generous estimate of extinction coefficient or better aligned transition dipoles between donor and acceptor QDs. Based on the kinetic Monte Carlo simulations described below, we compute that the required $R_o$ to support the measured migration length of 35 nm within the 5 ns TRUSTED measurement window is ~ 15 nm. Because of the $1/r^6$ scaling of FRET, the discrepancy in $R_o$ result in 100-1000 times difference in exciton diffusivity. It is common that FRET theory underestimates the measured exciton transport rate in quantum dot solids(*9, 11, 16, 17*).

## 6. Kinetic Monte Carlo simulation setup

We used Kinetic Monte Carlo (KMC) simulations to examine whether adjusting $R_o$ alone can reconcile all measurements of energy transfer.

In a hexagonally packed QD grid, each QD is assigned an energy randomly sampled from a distribution corresponding to the peak emission distribution from the single-particle measurements (**Fig. 1D,E**). The FRET rate from a donor QD to every other possible acceptor QD is calculated from Eq. **S4 and S5** above. The spectral overlap component $\int \sigma_A(\lambda)\overline{F_D(\lambda)}\lambda^4 d\lambda$ can be rewritten in terms of energy $\varepsilon$ as:

$$\int \frac{\sigma_A(\varepsilon)F_D(\varepsilon)}{\varepsilon^4} d\varepsilon$$

*Eq. S6*

We used values of $\Phi_D$ and $\eta$ according to **Section 5**. Dipole orientations of all QDs are assumed to have a random orientation, so $\kappa^2 = 2/3$ for all donor and acceptor pairs. The spectral overlap integral is calculated using

$$\sigma_A(\varepsilon) = \sigma_A(\varepsilon_A) \exp\left[-\frac{(\varepsilon - \varepsilon_A)^2}{2\sigma_h^2}\right]$$

*Eq. S7*

and

$$F_D(\varepsilon) = \frac{1}{\sigma_h\sqrt{2\pi}} \exp\left[-\frac{(\varepsilon - \varepsilon_D + \Delta_{SS})^2}{2\sigma_h^2}\right]$$

*Eq. S8*

where $\sigma_A(\varepsilon)$ is the acceptor absorption cross section and $F_D(\varepsilon)$ is the donor emission spectrum whose integral is normalized to one. Each QD possesses an intrinsic linewidth of $\sigma_h = 74$ meV (174 meV FWHM), the median of the single particle emission FWHM distribution determined by the single particle measurements. $\varepsilon_A$ and $\varepsilon_D$ are the randomly generated peak absorption energies of the acceptor and donor QDs, according to the distribution in **Fig. 1D,E**, with $\sigma_{ih} = 80$ meV. $\Delta_{ss}$ represents the Stokes shift.

The energy transfer rate between a donor and an acceptor is determined by **Eq. 2 in the main text.** The time for the transfer event is the inverse of the summation of all rates. An acceptor site for a given hop is selected by a random distribution weighted by the rate of each pair. In other words, pairwise interactions with faster rates are more likely to be chosen.

In the simulations below, we aim to examine each $r$ dependence in **Eq. 2 of main text** separately by matching the measured exciton diffusion extent within the TRUSTED window, $35 \pm 6$ nm, with simulation and comparing the resulting mean exciton energy relaxation to the experimental TRES. Specifically, we changed the $r$ dependence in the rate equation and adjusted the $\sigma_A(\varepsilon_A)$ as a linear coefficient of energy transfer rate to match the 35 nm exciton diffusion extent, while keeping other parameters untouched. **Fig. S15** shows the resulting decay of mean exciton energy versus the experimental TRES data. Clearly, $r^{-6}$ dependency produces an energy decay that is too fast while the $r^{-2}$ regime predicts a decay that is too slow. This strongly suggests that there are multiple exciton transport pathways with different $r$ dependencies in QDSL.

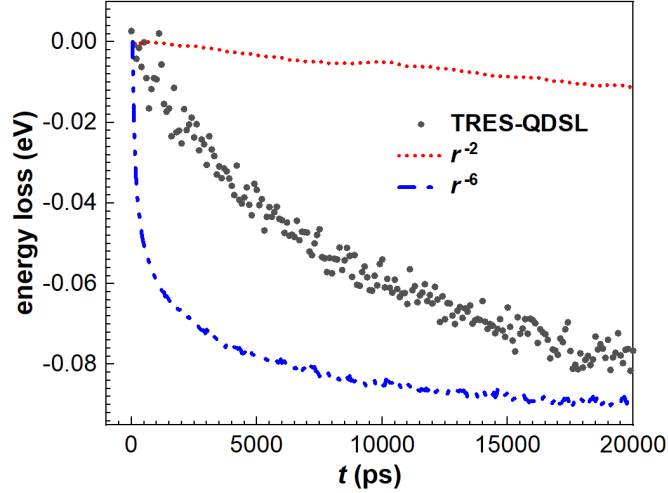

**Figure S15. Comparison of far-field and near-field coupling terms in isolation attempting to reconcile the TRUSTED and TRES experimental results.**

To evaluate the relative amplitude of the three regimes of dipole-dipole coupling, we used **Eq. 2 of main text** to calculate energy transfer rate in the simulation. We gradually increase $c_{\text{far}}$ and adjust $\sigma_A(\varepsilon_A)$ accordingly so that a matching between TRUSTED and TRES may be achieved simultaneously. In the example below, when $c_{\text{far}} \sim 27$, both TRUSTED and TRES experimental data can be reproduced by the simulation.

**Contribution of near-field and far-field components in the model**

As explained in the main text, when the energy transfer rate is described as

$$k = \frac{1}{\tau_D} \frac{R_{o,}^6}{r^6} (1 + c_{\text{intermed}}(2\pi r/\lambda)^2 + c_{\text{far}}(2\pi r/\lambda)^4) \exp(-\alpha r)$$

*Eq. S9*

with $c_{\text{far}} \sim 27$, and the average $R_o = 7.6$ nm, both the 35 nm exciton diffusion extent within the TRUSTED window and the energy relaxation dynamics from TRES can be reconciled.

To investigate the contribution of near-field coupling, we performed simulations using strictly near-field interactions with

$$k = \frac{1}{\tau_D} \frac{1}{r^6} (R_o^6) \exp(-\alpha r).$$

*Eq. S10*

In other words, as a control, the far-field component was removed, and all other parameters were retained. With **Eq. S10** as the energy transfer rate, the simulation yields a 3.7 nm exciton diffusion extent within the TRUSTED window. Thus, the far-field coupling is essential to increase the exciton diffusion extent to the experimental value of 35 nm.

In addition, we analytically calculated the percentage of near-, intermediate-, and far-field coupling mediated hops. To do this, we multiplied the three terms individually by the number of acceptors, which is proportional to $r^2$, and performed integration along $r$ out to infinity. This calculation shows that 6% of the rate is contributed by the intermediate- and far-field coupling terms. To identify the relative contributions to $c_{\text{far}}$ we note that in the isotropic limit, it includes a factor of 1/3 due to orientational averaging of the projection of donor and acceptor transition dipole moments. While this orientational factor has already been accounted for, it remains to account for the following: we estimate using Gaussian spectral profiles that the far-field overlap integral $\int \sigma_A(\varepsilon) F_D(\varepsilon) d\varepsilon$ is 17× greater than the corresponding weighted integral for the near-field term embodied in $R_o$ in **Eq. S6**. As a result, we find that the relative strength of the far field term requires a remaining factor of 27/17≈5 to adequately model the experimental data. This estimate is only approximate because **Eq. S9** accounts for the scaling with $\lambda$ (or $\varepsilon$) in the near-field term without the careful integration that is otherwise carried out in **Eq. S6**. The possible contributions to this factor, irrespective of its absolute amount, are discussed in the main text. We also note that the full relative strength of far-field to near-field contributions to the energy transfer rate is smaller than 1 part in $10^6$, as one must also account for the factor of $(2\pi r/\lambda)^4$ that we used to non-dimensionalize $c_{\text{far}}$.

## 7. Electrodynamics simulation demonstrate enhanced far-field over near-field strength

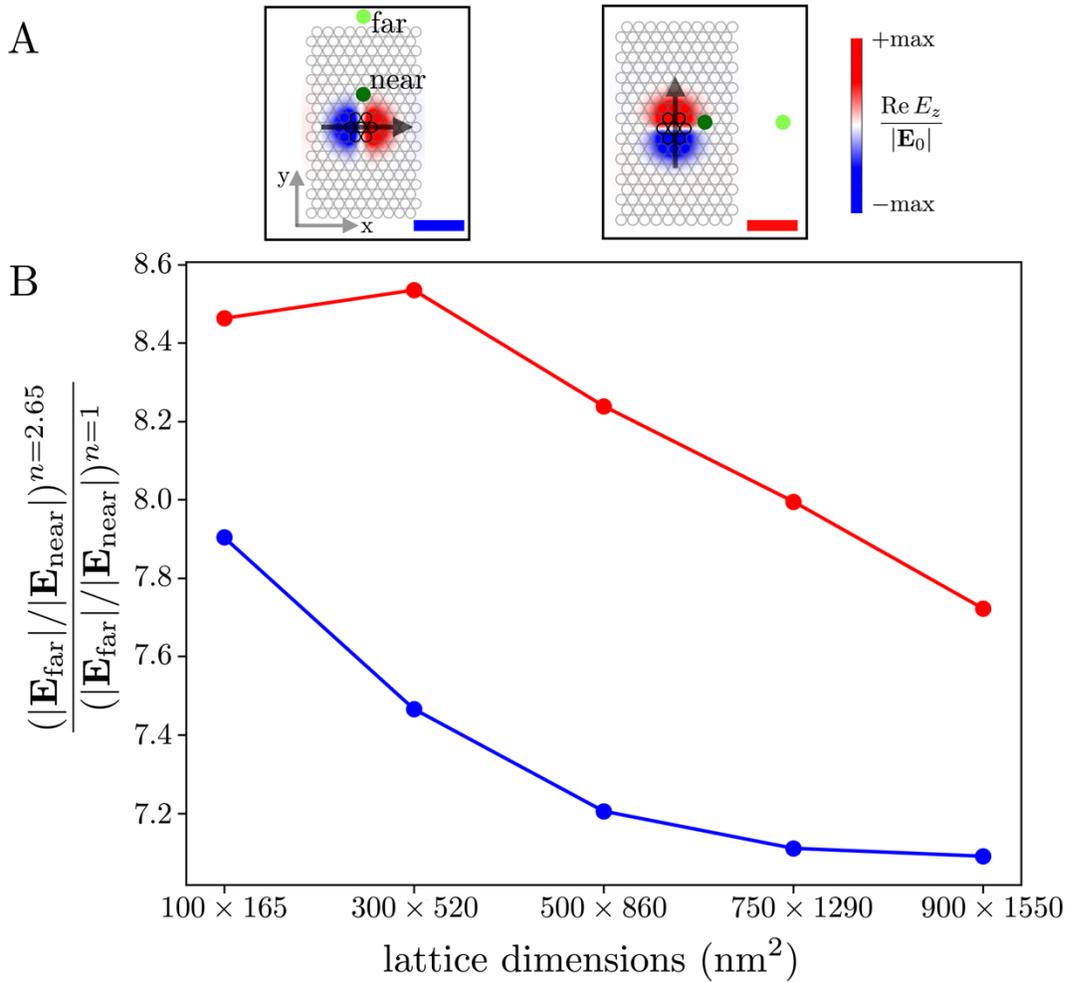

**Figure S16**: **Comparison of far-field to near-field electric field strength computed in hexagonally packed CdSe QD superlattices with background refractive indices of $n=2.65$ and $n=1$ as a function of overall lattice size.** (A) Representation of the two studied localized QD emission polarization directions and points of observation: (left) horizontal emission polarization and vertical observation, (right) vertical emission polarization and horizontal observation. The out-of-plane component of the localized QD emission electric field ($E_z$) is displayed in each case. (B) Computed far-field to near-field electric field magnitude ratios corresponding by color to each case in (A) as a function of overall lattice size with background refractive indices of $n=2.65$ and $n=1$. In each case the far- and near-field observation points are 7 μm and 30 nm away from the QD excitation region, and the excitation frequency is $2.98 \times 10^{15}$ rad·sec$^{-1}$.

**Figure S16** illustrates the ratio of the far- to near-field electric field magnitude as a function of hexagonally packed "CdSe QD" superlattice size for background refractive indices of $n=2.65$ and $n=1$. Within the coupled dipole approximation(55–57), the QD polarization in each lattice point is represented as a point electric dipole that can be polarized by an applied excitation electric field ($E_0$) as well as by the fully retarded electric dipole fields of all surrounding lattice points. The polarizability of each effective CdSe QD is taken from the $l$=1 electric Mie coefficient corresponding to a 9 nm diameter sphere using the frequency-dependent bulk optical constants from Ref. (*58*). The center-to-center distance between adjacent QD spheres is 10 nm. In each case considered, only a single hexagonal QD cluster positioned at the lattice center is excited at frequency $2.98 \times 10^{15}$ rad·sec$^{-1}$ by windowing an applied plane wave electric field to that cluster (shown as the black central hexagon in panel A). The QD emission and excitation frequencies and polarizations are the same as the applied field. Two directions are considered for the localized QD excitation/emission polarization (horizontal and vertical) with corresponding in-plane observation points oriented perpendicular to the polarization direction to maximize overlap with the far-field lobes of the excited QD dipoles as shown in panel A. Flooding the background with the CdSe refractive index of $n$=2.65 (at frequency $2.98 \times 10^{15}$ rad·sec$^{-1}$) is chosen in the computed finite lattice systems to better approximate the wavelength contraction effects that would occur in our experimentally studied CdSe QD superlattices. Panel B compares the ratio of the far-field magnitude (computed 7 μm ≈ 10·λ away from the excited QD region) to the near-field magnitude (computed 30 nm ≪ λ away from the excited QD region) in the higher index ($n$=2.65) medium compared to vacuum ($n$=1), where λ is the in-medium emission wavelength. A 7-8fold enhancement of the far-field over the near-field is evident across all lattice dimensions studied; similar enhancement would remain even when extrapolating the lattice to the larger dimensions relevant to experiment. This latter effect is already suggested by comparing the left-hand system configuration to the right-hand one in panel A, since on the left there is more polarizable lattice between the source and far-field observation point, and the field ratios in B appear to already asymptote near a value of 7. Taken together, these computational results include both the waveguiding effects occurring in finite CdSe QD superlattices as well as the associated wavelength contraction effects experienced by the electromagnetic signals relaying energy between donor and acceptor points within the lattice, and show a lattice-induced enhancement in the far- to near-field ratio in all cases studied numerically.

## 8. Far field coupling inclusion into previous literature results

We also performed KMC simulation to test our model on literature results. We chose to test our model on the QD solid measured by Akselrod *et al*(*9*) because they examined the decay of mean exciton energy by TRES as well as spatiotemporal exciton migration via transient PL microscopy. However, no direct connection was established between the two experiments. To do the KMC simulation, we use parameters corresponding to their "$d_1$" QDs, $\sigma_h$ = 30 meV, $\sigma_{ih}$ = 23 meV, $\Delta_{ss}$ = 38 meV. Similarly, $\sigma_A$ is adjusted to match the experimental transient PL microscopy or TRES results. Here, $c_{far}$ ~ 35 is needed to reconcile their reported exciton diffusion length and energy relaxation rate. Furthermore, in this case, $R_o$ becomes ~ 5 nm as opposed to 11 nm obtained with FRET alone,(*9*) well within the reasonable range of estimation based on steady state spectral measurements.(*22*). This example demonstrates that by tuning the inhomogeneity of the QD linewidth, one may control the ratio between near-field and far-field interactions."

## 9. Tuning the far-field contribution to exciton transport via QD energy inhomogeneity

Based on the Kinetic Monte Carlo simulation that reproduces both the TRUSTED and TRES data of the 5% Te doped CdSe/CdS QDSL (labeled "control simulation"), we repeated the simulation while systematically tuning the intrinsic and inhomogeneous linewidth of the QD absorption/emission spectrum, all the while maintaining the same total linewidth. We report the percentage of exciton hops whose displacement is greater than 40 nm (POH40) over the total number of exciton hops at both 0 ns and 20 ns after photoexcitation.

First, with a FRET-only model, the percentage of hops that exceed 40 nm is extremely small, on the order of 0.1%, as is shown in the first column of **Table S2** labeled "FRET only."

Second, as the intrinsic linewidth is decreased and the inhomogeneous linewidth is increased from columns "Less inhomogeneous" to "Control simulation" to "More inhomogeneous," while POH40 at 0 ns is similar in each case, POH40 differs substantially at 20 ns. For the less inhomogeneous system, POH40 increases to 9.7%, in contrast to the decrease to 4.2% exhibited by the more inhomogeneous system. This difference is in part due to the complex interplay of three energetic parameters that determine energy resonance in the dipole-dipole coupling theory: Stokes shift, inhomogeneous linewidth, and intrinsic linewidth[10]. This example demonstrates that by tuning the inhomogeneity of the QD linewidth, one may control the ratio between near-field and far-field interactions.

**Table S2. Tuning QD energy inhomogeneity can alter the effect of far-field interactions.**

|  | FRET only | Less inhomogeneous | Control simulation | More inhomogeneous |
|---|---|---|---|---|
| Intrinsic linewidth (meV) | 74 | 104.8 | 74 | 30 |
| Inhomogeneous linewidth (meV) | 80 | 30 | 80 | 104.8 |
| Percentage of far-field hops@0ns | 0.13% | 5.0% | 4.8% | 5.6% |
| Percentage of far-field hops@20ns | 0.06% | 9.7% | 4.8% | 4.2% |
| Diffusion length within 4.8 ns | 5.6 nm | 46 nm | 35 nm | 30 nm |